\documentclass[twocolumn,floatfix,altaffilletter,superscriptaddress,preprintnumbers,
               tightenlines,showpacs,showkeys,nofootinbib]{revtex4-1}
\usepackage[utf8]{inputenc}
\usepackage[colorlinks=true,citecolor=blue,linkcolor=blue]{hyperref}
\usepackage[normalem]{ulem}
\usepackage{amsmath,amssymb}
\usepackage{epsfig}
\usepackage{graphicx}               % Standard graphics package
\usepackage{url}
\usepackage{color}
\usepackage{slashed}
\usepackage{multirow}
\usepackage{placeins}
\usepackage[dvipsnames]{xcolor}
\usepackage{epstopdf}
\usepackage{soul}
\usepackage{tikz}
\usepackage[capitalise, english]{cleveref}
\usepackage{siunitx}
\usepackage{xspace}
\usetikzlibrary{trees}
\usetikzlibrary{decorations.pathmorphing}
\usetikzlibrary{decorations.markings}

\definecolor{jblue}{RGB}{20,50,100}
\definecolor{npurple}{RGB} {153, 51, 204}
\definecolor{wred}{RGB}{217,0,56}
\definecolor{white}{RGB}{255,255,255}

\definecolor{korange}{RGB}{235, 80,  43}
\definecolor{korange2}{RGB}{245, 100,  63}
\definecolor{kyelloworange}{RGB}{255, 210,  110}
\definecolor{kyelloworange2}{RGB}{240, 170,  90}
\definecolor{kred}{RGB}{204,  102, 153}
\definecolor{kpurple}{RGB}{153,  61, 190}
\definecolor{kpurplelight}{RGB}{213,  161, 230}

\DeclareSIUnit\year{yr}
\DeclareSIUnit\pc{pc}
\DeclareSIUnit\ergs{ergs}
\DeclareSIUnit\msun{\ensuremath{M_\odot}}
\allowdisplaybreaks

\newcommand{\ev}[1]{\ensuremath{\left\langle #1 %
                     \right\rangle}} % Expectation value
\renewcommand{\vec}[1]{{\mathbf{#1}}}
\newcommand{\iso}[2]{{\ensuremath{{}^{#2}}\ensuremath{\rm #1}}}

\providecommand{\abs}[1]{\lvert#1\rvert}

\newcommand{\altp}{\ensuremath{\widetilde{\alpha}^\prime}\xspace}
\newcommand{\btp}{\ensuremath{\widetilde{\beta}^\prime}\xspace}
\newcommand{\alp}{\ensuremath{\alpha^\prime}\xspace}
\newcommand{\mmed}{\ensuremath{m_{\rm med}}\xspace}
\newcommand{\vesc}{\ensuremath{v_{\rm esc}}}
\newcommand{\qchi}{\ensuremath{q_\chi}}

\makeatletter
\providecommand*{\diff}%
  {\@ifnextchar^{\DIfF}{\DIfF^{}}}
\def\DIfF^#1{%
  \mathop{\mathrm{\mathstrut d}}%
    \nolimits^{#1}\gobblespace}
\def\gobblespace{%
  \futurelet\diffarg\opspace}
\def\opspace{%
  \let\DiffSpace\!%
  \ifx\diffarg(%
    \let\DiffSpace\relax
  \else
    \ifx\diffarg[%
      \let\DiffSpace\relax
    \else
        \ifx\diffarg\{%
        \let\DiffSpace\relax
      \fi\fi\fi\DiffSpace}


\begin{document}

%=============================================================================
\title{Cuckoo's Eggs in Neutron Stars: Can LIGO Hear Chirps from the Dark Sector?}

\author{Joachim Kopp}
\email{jkopp@uni-mainz.de}
\affiliation{PRISMA Cluster of Excellence and
             Mainz Institute for Theoretical Physics,
             Johannes Gutenberg-Universit\"{a}t Mainz, 55099 Mainz, Germany}
\affiliation{Theoretical Physics Department, CERN, 1211 Geneva, Switzerland}

\author{Ranjan Laha} 
\email{ranjalah@uni-mainz.de}
\author{Toby Opferkuch}
\email{opferkuch@uni-mainz.de}
\author{William Shepherd}
\email{shepherd@uni-mainz.de}
\affiliation{PRISMA Cluster of Excellence and
             Mainz Institute for Theoretical Physics,
             Johannes Gutenberg-Universit\"{a}t Mainz, 55099 Mainz, Germany}

\date{\today}

\preprint{MITP/18-058}
%=============================================================================

\begin{abstract}
We explore in detail the possibility that gravitational wave signals from
binary inspirals are affected by a new force that couples only to dark matter
particles.  We discuss the impact of both the new force acting between the
binary partners as well as radiation of the force carrier.  We
identify numerous constraints on any such scenario, ultimately concluding that
observable effects on the dynamics of binary inspirals due to such a force are
not possible if the dark matter is accrued during ordinary stellar evolution.
Constraints arise from the requirement that the astronomical body be able to
collect and bind at small enough radius an adequate number of dark matter
particles, from the requirement that the particles thus collected remain bound
to neutron stars in the presence of another neutron star, and from the
requirement that the theory allows old neutron stars to exist and retain their
charge.  Thus, we show that any deviation from the predictions
of general relativity observed in binary inspirals must be due either to the
material properties of the inspiraling objects themselves, such as a tidal
deformability, to a true fifth force coupled to baryons, or to a
non-standard production mechanism for the dark matter cores of neutron stars.
Viable scenarios of the latter type include production of dark matter
in exotic neutron decays, or the formation of compact dark matter objects in
the early Universe that later seed star formation or are captured by stars.
\end{abstract}

\maketitle

%-----------------------------------------------------------------------------
\section{Introduction}
%-----------------------------------------------------------------------------

The dawn of gravitational wave (GW) astronomy, commencing with the measurement
of GWs produced by coalescing binary black holes (BHs), promises an entirely
new era of observational astrophysics \cite{Abbott:2016blz,
TheLIGOScientific:2016pea, Abbott:2016nmj}. Recently, coalescing neutron star
 binaries have also been observed for the first time
\cite{TheLIGOScientific:2017qsa}. Such events are treasure troves of new data
for astrophysics and astroparticle physics.  They not only feature an
associated electromagnetic signal \cite{Coulter:2017wya, Monitor:2017mdv, GBM:2017lvd}, but also produce gravitational waveforms that are potentially observable
over time periods of order 20~minutes and at length scales which are different
from observations of binary pulsars \cite{Stairs:2003eg}.

The ability to track
the neutron star binary inspiral over an extended period of time provides
increased sensitivity to modifications of the post-Newtonian phase of the
inspiral compared to inspirals of $\mathcal{O}(30 M_\odot)$ black holes.  These
deviations can be broadly categorized as stemming either from \emph{(i)} a
modification of gravity \cite{Belenchia:2016bvb, Cai:2017cbj, Sagunski:2017nzb,
  Burrage:2017qrf, Nojiri:2017hai, Dima:2017pwp, Berti:2018cxi, Gong:2018cgj,
Berti:2018vdi, Abedi:2018npz}, \emph{(ii)} a ``fifth force'' mediated by
particles such as axions \cite{Hook:2017psm} or ultra-light dark gauge bosons
\cite{Croon:2017zcu}, or from \emph{(iii)} a new long-range force coupled to
dark matter (DM) particles accreted inside neutron stars \cite{Ellis:2017jgp,
Nelson:2018xtr, Ellis:2018bkr}. Stringent constraints exist on a new fifth
force coupled to baryons, see for instance Refs.\ \cite{Adelberger:2009zz,
Salumbides:2013dua, Salumbides:2013aga, Berge:2017ovy, Fayet:2017pdp} and
references therein, which have motivated a particular focus on the possibility of a new
long-range force, mediated by ultra-light bosons, that couples only to dark
matter or other cosmologically stable hidden sector particles residing inside
neutron stars.  Such scenarios have been studied extensively using probes
other than gravitational waves, in particular the properties (and the mere
existence) of old neutron stars and white dwarfs~\cite{Goldman:1989nd,
  Kouvaris:2007ay, Sandin:2008db, Kouvaris:2010vv, deLavallaz:2010wp,
  Ciarcelluti:2010ji, McDermott:2011jp, Kouvaris:2011fi, Leung:2011zz,
  Kouvaris:2011gb, Guver:2012ba, Li:2012qf, Bramante:2013hn, Bell:2013xk,
Leung:2013pra, Goldman:2013qla, Bramante:2013nma, Zheng:2014fya,
Mukhopadhyay:2016dsg, Baryakhtar:2017dbj, Raj:2017wrv, Cermeno:2017xwb,
Rezaei:2018iyu, Kouvaris:2018wnh, Wang:2018wak, Chen:2018ohx,Chen:2018ohx} as well as the dynamics of binary pulsar PSR~1913+16 (Hulse--Taylor binary) \cite{Krause:1994ar,
Mohanty:1994yi}. We note that dark sector particles experiencing a strong
long-range force can at most contribute a small fraction to the total DM
density of the Universe due to tight constraints on DM self-interactions
\cite{Kaplinghat:2015aga, Tulin:2017ara, Braaten:2018xuw}. 

Complementing these results, we will show that DM accumulation in neutron stars
throughout the course of ordinary stellar evolution can never be efficient
enough for the new force to lead to observable deviations in gravitational wave
signals from binary inspirals.  Detectable, per cent-level deviations require
$\gtrsim 10\%$ of the neutron star mass to be made up of dark sector
particles. Exploring the constraints on the amount of such particles that
a neutron star can contain, we will conclude that only exotic
neutron decays or the formation of dark stars can result in such large abundances.

We begin in \cref{sec:inspiral_signal} by reviewing fifth-force effects in
neutron star binary inspirals, focusing on the required region of parameter
space to observe deviations from general relativity in LIGO/VIRGO data. Of
course, our results are easily generalized to other gravitational wave
detectors such as ALIA, BBO, DECIGO, the Einstein Telescope, Geo-600, KAGRA,
2, TAMA, etc.~\cite{Pitkin:2011yk,Will:2014kxa}.
We then derive constraints on these forces, distinguishing
between repulsive forces (\cref{sec:repulsive}) and attractive ones
(\cref{sec:attractive}). We conclude in \cref{sec:conclusions}.  Note that we
use natural units throughout the text.

%-----------------------------------------------------------------------------
\section{Inspirals and fifth-forces}
\label{sec:inspiral_signal}
%-----------------------------------------------------------------------------

Before turning to the constraints on the interactions of any exotic hidden sector
particle, we first illustrate the necessary conditions for an observable new physics
signal at LIGO/VIRGO. In doing so, we closely follow Refs.~\cite{Sagunski:2017nzb,
Croon:2017zcu}.  To remain as model-independent as possible, we parameterize the
effect of the new dark force using a generic Yukawa potential. We furthermore
assume that the new force couples only to dark sector particles.  In order for
it to act over sufficiently long distances of order 100~km, the mediator must
be ultra-light ($\lesssim 10^{-12}$~eV), and the dark sector particles must
neither significantly screen the new force nor efficiently self-annihilate. The
most straightforward way of realizing the second constraint is asymmetric DM
\cite{Petraki:2013wwa, Zurek:2013wia}. 

Let us consider a binary neutron star system, and let us assume that each of
the binary partners contains a population of dark sector particles. These
populations will affect the inspiral dynamics in two distinct ways: first, the
exotic force acting between them will affect the time evolution of the distance
between the neutron stars, their orbital frequency, and the time of the merger.
Second, radiation of the new light force carriers provides an extra energy loss
mechanism.

\subsection{Effects of a new Yukawa force}
\label{sec:dark-force}
%-----------------------------------------

Even in the presence of a new dark sector Yukawa force, the system's
gravitational wave emission will follow the predictions of general relativity
as long as the distance between the binary partners is significantly larger
than the range of that force.  However, once their distance drops below that
range and the exponential suppression of the Yukawa potential is lifted,
observable deviations may occur.  The assumption of a Yukawa force is necessary,
as the potential effects of a new infinite range force are degenerate with a
shift of the neutron star masses. The new force, which can be either attractive
or repulsive, results in a modification of the chirp mass $\mathcal{M}_c \equiv
\mu^{3/5} (M_1 + M_2)^{2/5}$, where $\mu=M_1 M_2/(M_1 + M_2)$ is the reduced
mass of the inspiraling objects with individual masses $M_1$ and $M_2$.
Assuming point-like neutron stars, we can write the magnitude of the force
between the two bodies as
\begin{align}
  |\vec{F}| &= \frac{G_N M_1 M_2}{\Delta^2}
               \left[1 + \altp e^{-\mmed \Delta}(1+\mmed\Delta)\right]\,,
\end{align}
where $\Delta$ is their spatial separation and \mmed is the mass of the
particle sourcing the Yukawa potential.  Here $\altp$ parameterizes the size of
the new force relative to gravity in the regime $\mmed\Delta\ll1$:
\begin{align}
  \label{eq:altp_definition}
  \altp \equiv \pm \frac{\alp Q_1 Q_2}{G_N M_1 M_2}\,.
\end{align}
The sign in this expression determines whether the force is attractive
(positive sign) or repulsive (negative sign). $\alp$ is the
analog of the fine structure constant for the new force, and $Q_i = q_\chi
N_{\chi i}$ are the total charges of the inspiraling objects under the new
force, which depend on the charge $q_\chi$ of a single DM particle and the
numbers $N_{\chi i}$ of captured DM particles.  Given that the orbits of the
binary system will have circularized by the time the gravitational wave signal
becomes observable \cite{Peters:1964zz}, the orbital frequency of the system is
given simply by a modified Kepler's law,\footnote{Note the factor $(1 + \mmed \Delta)$ was neglected in ref.~\cite{Croon:2017zcu}, hence \cref{eq:modified_kepler} and those that follow disagree with this reference.}
\begin{align}
  \label{eq:modified_kepler}
  \omega^2 &= \frac{G_N (M_1 + M_2)}{\Delta^3}
              \left[1+ \altp e^{-\mmed \Delta}(1+ \mmed \Delta)\right]\,.
\end{align}
The frequency of the gravitational wave signal is $f_\text{GW} = \omega/\pi$;
therefore if LIGO sensitivity begins at
$\mathcal{O}(\SI{10}{\Hz})$ \cite{TheLIGOScientific:2017qsa}, \cref{eq:modified_kepler}
tells us that a binary system consisting of two $1.25\, M_\odot$ neutron stars
enters the sensitivity band at a spatial separation of $\mathcal{O}(\SI{700}{\km})$
(assuming that gravity is the dominant force at these length
scales, i.e.\ $\abs{\altp}\ll1$). A typical neutron star radius being \SI{10}{\km}
indicates that a Yukawa force detectable by LIGO must have range $\mmed^{-1}
\simeq \mathcal{O}(\SI{20}{}-\SI{750}{\km})$, i.e.\ $\mmed \simeq
\SI{E-11}{}$--$\SI{3E-13}{eV}$.

To determine the inspiral dynamics we require the total energy of the system,
\begin{align}
  E_\text{tot} &= -\frac{G_N \mu (M_1 + M_2)}{\Delta} \left(1+\altp e^{-\mmed \Delta}\right)
                 + \frac{1}{2}\mu \Delta^2 \omega^2\,.
  \label{eq:total-energy}
\end{align}
The last term is the kinetic energy of the stars, neglecting their spin and internal
structure. The power radiated via gravitational waves is
\cite{Abbott:2016bqf, Mathur:2016cox}
\begin{align}
  \frac{\diff E_\text{GW}}{\diff t} &= \frac{32}{5}\, G_N \mu^2 \Delta^4 \omega^6\,.
\end{align}
Equating $-\diff E_\text{GW} / \diff t$ to the time derivative of
\cref{eq:total-energy}, we obtain the rate of change of the orbital frequency,
\begin{align}
  \label{eq:frequency_derivative}
  \frac{\diff \omega}{\diff t} &= -\frac{32}{5}\,G_N \mu \Delta^2 \omega^5
                                   g(\altp,\mmed,\Delta)\,,
\end{align}
where
\begin{align}
  g &= -\frac{3+\altp e^{-\mmed \Delta}\left(3+\mmed \Delta(3+\mmed \Delta)\right)}
             {1+\altp e^{-\mmed \Delta}\left(1+\mmed \Delta(1-\mmed \Delta)\right)}\,.
\end{align}
In the massless mediator limit, this reduces to $g=-3$, and the classical gravity-only
result is recovered \cite{Abbott:2016bqf}. In this limit the effect of the infinite range
fifth-force results merely in a modification of the apparent strength of gravity,
parameterized by the replacement $G_N \to G_N (1 + \altp)$ in \cref{eq:modified_kepler}.
\Cref{eq:modified_kepler} can then be used to eliminate $\Delta$ from
\cref{eq:frequency_derivative}, yielding
\begin{align}
  \frac{\diff \omega}{\diff t} &= \frac{96}{5} (G_N \mathcal{M}_c)^{5/3}
                                  (1+\altp)^{2/3} \omega^{11/3}\,.
  \label{eq:frequency_derivative-2}
\end{align}
We see that an infinite range Coulomb-like force results in a modified chirp mass
\begin{align}
  \widetilde{\mathcal{M}}_c \equiv \mathcal{M}_c \left(1+\altp\right)^{2/5}\,.
\end{align}
The system is thus indistinguishable from a purely gravitationally interacting
system with a different chirp mass. For non-zero \mmed,
\cref{eq:modified_kepler,eq:frequency_derivative} must be solved numerically,
as there is no analytic solution for $\Delta(\omega)$ in \cref{eq:modified_kepler}. 

\begin{figure}
  \centering
  \includegraphics[width=\columnwidth]{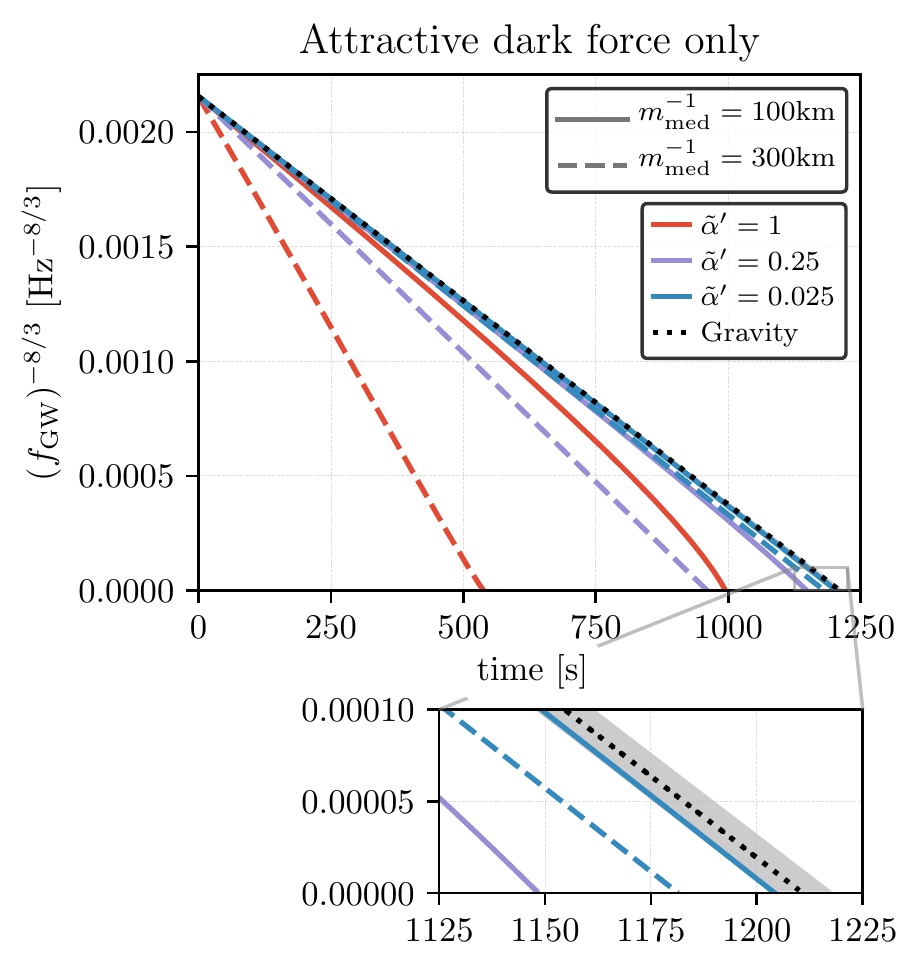}
  \caption{Gravitational wave frequency $f_\text{GW}^{-8/3}(t)$ as a function of
    time for a neutron star binary with various values of the
    fifth-force parameters \altp and \mmed. In particular, we show results for
    $\mmed^{-1}=\{\SI{100}{},\SI{300}{}\}\SI{} {\km}$ (solid, dashed lines)
    and three choices of $\altp=\{1,0.25,0.025\}$ (red, purple, and blue colored
    lines). We consider here the limit that the system loses energy only by
    gravitational wave emission, but not by radiation of light mediator
    particles. This situation is realized approximately if the binary partners
    have equal dark charge-over-mass ratios.
    The gray band indicates the typical error on reconstructing the chirp mass,
    which we conservatively take to be 0.4\% based on ref.~\cite{TheLIGOScientific:2017qsa}.
    The boundary condition of the differential equation is the threshold
    sensitivity of LIGO, $f_\text{GW}(0)=\SI{10}{\Hz}$ \cite{Martynov:2016fzi}.
    The neutron star masses are $M_1=M_2=1.25 M_\odot$.
  }
  \label{fig:Frequency_deviation}
\end{figure}

In \cref{fig:Frequency_deviation}, the gravitational wave frequency is shown
as a function of time for three choices of $\altp=\{1,0.25,0.025\}$, and for
two different values of $\mmed^{-1} = \{\SI{100}{},\SI{300}{}\}~\SI{}{\km}$.
On the vertical axis, we plot $f_\text{GW}^{-8/3}$ rather than just
$f_\text{GW}$ because this choice yields straight lines in the gravity only
scenario (dotted black line in \cref{fig:Frequency_deviation}). In the pure
gravity scenario, the unknown source distance causes the largest uncertainty
in reconstructing the chirp mass.  This uncertainty was estimated as
$\sim$0.4\% in the discovery paper of
GW170817~\cite{TheLIGOScientific:2017qsa}.  However, the chirp mass in the
detector frame has a much smaller uncertainty, $\lesssim$ 0.067\%.  As a
conservative choice, we show the larger uncertainty, i.e, 0.4\%, as the grey
band in \cref{fig:Frequency_deviation} and we use this value throughout the
text.  A more precise knowledge of the source distance will improve the
constraints on new physics considerably. For $\mmed^{-1}=\SI{300}{\km}$ and
large values of \altp, we see that ${\diff\omega}/{\diff t}$ is a still
constant, but clearly different to the gravity only solution, for the entire
frequency range in which LIGO is sensitive. In other words, the Yukawa force
looks like a Coulomb force to LIGO, and we would simply reconstruct the event
with an alternative, \emph{incorrect}, value of the chirp mass.  There is thus
no sensitivity to the new fifth-force for these parameters from GW signals
alone; an electromagnetic signal associated with a GW event with reconstructed
masses above the Chandrasekhar limit would indicate a modification induced by
a dark force.

On the other hand,
larger mediator masses and smaller fifth-force couplings lead to deviations
that occur part-way through the time domain of the LIGO sensitivity band. These
scenarios would be observed as prematurely coalescing binaries. Even for $\altp=0.025$ and
$\mmed^{-1}=\SI{100}{\km}$, the coalescence time is shifted by $\sim \SI{6}{\s}$
relative to that predicted by gravity alone (with the chirp mass determined from
the first part of the wave form). In this work, we do not
consider higher order corrections to the gravitational wave emission. However,
we note that these
corrections can potentially break the degeneracy between the
effects of the dark force and
gravity in the case where the Yukawa potential of the dark force is
unsuppressed
throughout the observation window of the waveform \cite{Alexander:2018qzg}.
Finally, a more accurate
observable can be defined by generating the predicted waveforms including these
corrections and comparing
the discrepancies with and without these corrections to the instrumental noise
curves.

The modification of Kepler's law, \cref{eq:modified_kepler}, changes the initial
separation of the neutron stars compared to the gravity-only case at the time
when the waveform enters the LIGO frequency band. This in turn can modify the
amplitude of the signal. For attractive (repulsive) dark forces one obtains
an increase (decrease) in the amplitude. More concretely, the gravitational wave
amplitude, $A_\text{GW}(t)$, is given by \cite{maggiore2008gravitational,
Croon:2017zcu}
\begin{align}
  A_\text{GW} &= \dfrac{4 G_N \mu}{d_L} \omega^2 \, \Delta^2 \,,
\end{align}
where $d_L$ is the luminosity distance of the GW source.  In the pure gravity
scenario, $A_\text{GW} \times d_L$ is $\SI{6.3E-22}{\mega\pc}$ at the time
when the GW emission enters the LIGO frequency band. In
the presence of a dark force with $|\altp| = 0.25$ and mediator mass
$\mmed^{-1}=\SI{300}{\km}$, the amplitude
at this time becomes \SI{6.65E-22}{\mega\pc} and \SI{5.96E-22}{\mega\pc}
for an attractive and repulsive dark force, respectively.

\subsection{Ultra-light Boson Radiation}
\label{sec:radiation}
%---------------------------------------

Another source of energy loss is radiation of the ultra-light mediator
particles \cite{Sagunski:2017nzb, Croon:2017zcu}. The leading contribution to
the power radiated via vector mediators in the multipole expansion is
\cite{Ross:2012fc,Krause:1994ar}
\begin{align}
  \label{eq:dEdt-dipole-V}
  \frac{\diff E^V_\text{dipole}}{\diff t}
    &= \frac{2}{3} \, \alp \mu^2 \gamma^2 \omega^4 \Delta^2 \\
    &\quad\times \operatorname{Re} \left\{ \sqrt{1-\left(\frac{\mmed}{\omega}\right)^2}
              \left[ 1 + \frac{1}{2} \left(\frac{\mmed}{\omega}\right)^ 2\right] \right\} \,, 
							\notag
\end{align}
where $\gamma \equiv Q_1/M_1 - Q_2/M_2$ is the difference in charge to mass
ratios of the binary partners. \Cref{eq:dEdt-dipole-V} thus illustrates that
dipole radiation is only possible if the two stars have different dark
charge-over-mass ratio.  For scalar mediators, the power of the dipole
radiation is \cite{Ross:2012fc,Krause:1994ar}   
\begin{align}
  \label{eq:dEdt-dipole-S}
  \frac{\diff E^S_\text{dipole}}{\diff t}
    &= \frac{1}{3} \, \alp \mu^2 \gamma^2 \omega^4 \Delta^2 \\
    &\quad\times \operatorname{Re}
      \left\{ \left[1 - \left(\frac{\mmed}{\omega}\right)^ 2\right]^{3/2} \right\} \,.
							 \notag
\end{align}
In what follows, we will focus on vector radiation as a concrete example. 

\begin{figure}
  \centering
  \includegraphics[width=\columnwidth]{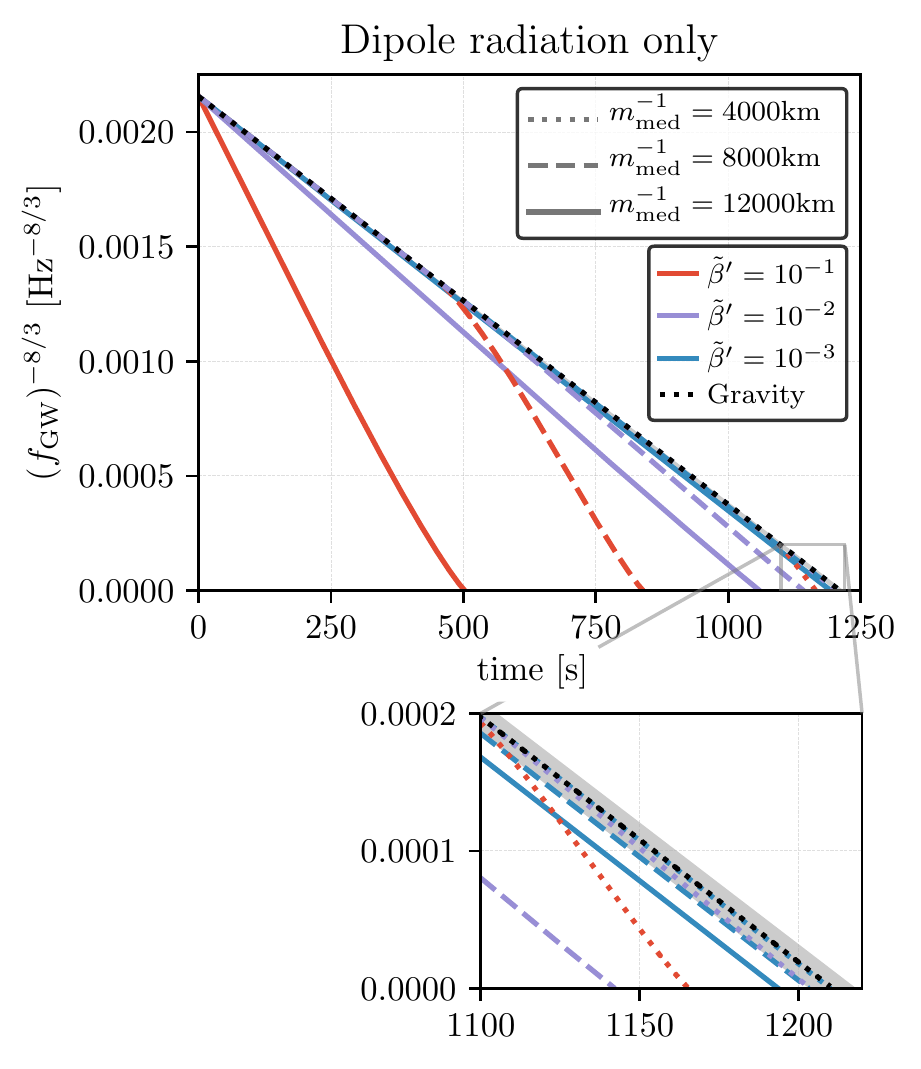}
  \caption{Gravitational wave frequency $f_\text{GW}^{-8/3}(t)$ as a function of
    time for a neutron star binary with various values of the fifth-force
    parameters \btp and \mmed.  We consider here the limit that the system
    loses energy via dipole radiation and gravitational wave emission, but
    that the latter is unaffected by a dark force acting between the binary
    partners. This situation is realized if one of the stars does not contain
    any dark sector particles.  We show results for
    $\mmed^{-1}=\{\SI{4000}{},\SI{8000}{},\SI{12000}\}\SI{} {\km}$ (dotted,
    dashed, solid lines) and three choices of $\btp = \{10^{-1}, 10^{-2},
    10^{-3}\}$ (red, purple, and blue colored lines).  The gray band indicates
    the typical error on reconstructing the chirp mass,
    which we take to be 0.4\% based on ref.~\cite{TheLIGOScientific:2017qsa}.
    The boundary condition of the differential equation is the threshold
    sensitivity of LIGO, $f_\text{GW}(0)=\SI{10}{\Hz}$ \cite{Martynov:2016fzi}.
    The neutron star masses are $M_1=M_2=1.25 M_\odot$.
  }
  \label{fig:Frequency_deviation_dipole}
\end{figure}

The energy loss to both gravitational wave and ultra-light boson radiation must
equal the orbital energy loss
\begin{align}
  \frac{\diff E_\text{tot}}{\diff t} &=
    - \left(\frac{\diff E_\text{GW}}{\diff t}
          + \frac{\diff E_\text{dipole}^V}{\diff t} \right) \,.
  \label{eq:equating_energy}
\end{align}
In \cref{fig:Frequency_deviation_dipole} we consider the limit where
the system loses energy to dipole radiation and to gravitational waves,
but gravitational wave emission is not affected by a dark force acting
between the binary partners. In other words, we set $\altp = 0$, but
$\btp \neq 0$ in
\cref{eq:equating_energy}.  This scenario corresponds to
the case of only one neutron star being appreciably charged.
We obtain
\begin{align}
  \frac{\diff \omega}{\diff t} &=
    \frac{96}{5} (G_N \mathcal{M}_c)^{5/3}\omega^{11/3}
  + \frac{1}{2} G_N (M_1+M_2) \btp \omega^3
                                                        \notag\\
  & \, \times
    \operatorname{Re} \left\{ \sqrt{1-\left(\frac{\mmed}{\omega}\right)^2}
                              \left[
                                1 + \frac{1}{2} \left(\frac{\mmed}{\omega} \right)^ 2
                              \right] \right\} \,,
  \label{eq:dipole_diffeqn}
\end{align}
where we have defined
\begin{align}
  \btp \equiv \frac{4 \alp \gamma^2 \mu^2}{G_N M_1 M_2} \,.
\end{align}
\btp parameterizes the magnitude of the radiation effect relative to gravity in
the same way as \altp from \cref{eq:altp_definition} parameterizes the effect
of a new force between the binary partners. There is a direct correspondence
\altp and \btp when the masses of the neutron stars are the same and we take
$Q_1=Q_2$ for \altp and $Q_2=0$ for \btp, i.e.  
\begin{align} 
  \left.\altp\right|_{Q_2=Q_1} &= \frac{\alp Q_1^2}{G_N M_1^2} 
                                = \left.\btp\right|_{Q_2=0} \,.
\end{align}

In \cref{fig:Frequency_deviation_dipole} we show solutions of
\cref{eq:dipole_diffeqn} varying both \mmed and \btp. The largest value of
$\mmed^{-1}$ shown, $\SI{12000}{\km}$, is chosen such that dipole radiation is
switched on while the waveform is still outside LIGO's frequency band. We see
that the effect of radiation is observable even in this case, in contrast to
the effect produced by a dark force with the same range, see
\cref{fig:Frequency_deviation}.  The reason is the different ways in which the
right hand sides of \cref{eq:frequency_derivative-2,eq:dipole_diffeqn} scale
with $\omega$.  For smaller values of $\mmed^{-1}$, the curves in
\cref{fig:Frequency_deviation} exhibit a pronounced kink around the point where
$\mmed \sim \omega$.  The kink is observable for $\btp \gtrsim 10^{-3}$, with
some dependence on $\mmed^{-1}$.  This means that the sensitivity to \btp
in the radiation-dominated scenario is about an order of magnitude better than
the sensitivity to \altp in the dark force-dominated scenario from
\cref{sec:dark-force}.  This difference arises because the two effects arise
at different order in an expansion in $\omega\Delta \sim 0.1$.

The upshot of this section is that for neutron star binaries in which both
stars carry a dark charge and have similar dark charge-to-mass ratio, LIGO may
be able to establish the existence of a new long-range force if $\altp \gtrsim
\mathcal{O}$(0.01) and $\SI{50}{km} \lesssim \mmed^{-1} \lesssim
\SI{300}{\km}$.  In binary systems with large dark charge-to-mass ratio
difference $\gamma$, observable effects from dipole radiation of the new force
carriers can be expected for $\btp \gtrsim \mathcal{O}(10^{-3})$ and
$\mmed^{-1} \gtrsim \mathcal{O}(\SI{4000}{\km})$. In this case, there is no
upper limit on the observable values of $\mmed^{-1}$. 

Of course, radiation
of dark force into higher multipoles is possible even for systems
that are symmetric in their charge-to-mass ratio. However, such radiation
is suppressed by additional powers of $\omega \Delta$, so we do not expect it
to dominate over the effect of the new long-range force between the binary
partners. Moreover, the effect of quadrupole radiation, like the effect of the
dark force, would be degenerate with the effect of gravity at large
$\mmed^{-1}$, making it unobservable in this regime independent of the
suppression by $\omega \Delta$.  At smaller $\mmed^{-1}$, however, radiation
into higher multipoles would lead to the same kinks as dipole radiation (see
\cref{fig:Frequency_deviation_dipole}), offering an extra handle for
establishing the radiation effect.  The kinks occur within the LIGO sensitivity
window for $\SI{4000}{\km} < \mmed^{-1} < \SI{12000}{\km}$. In the following sections we will
investigate the viability of the parameter values indicated above.

We note that the reach in $\mmed^{-1}$ may be significantly enhanced in the
future by combining LIGO/VIRGO measurements with results from observatories
which are sensitive to lower gravitational wave
frequencies\,\cite{Alexander:2018qzg}.  These observatories will be able to
observe neutron star binary systems long before the merger. If the separation
of the binary partners is larger than $\mmed^{-1}$ while the system is
observable by the lower frequency detectors, but below $\mmed^{-1}$ when it
enters the LIGO/VIRGO frequency band, the effect of the dark force can be
observed in a combined analysis.

To end this section we remark that while the above discussion has been in the
context of neutron stars these results are also applicable to black hole or
mixed binaries so long as the black holes can carry the appropriate charge.

\begin{table}[tb]
  \caption{The parameters of the benchmark neutron stars used in our
    numerical estimates. These parameter choices are representative of the recently
    observed neutron star binary merger \cite{TheLIGOScientific:2017qsa}.}
  \label{tab:neutron starparameters}
  \begin{ruledtabular}
  \begin{tabular}{ll}
    Neutron star parameter               & \\ \hline
    radius $R$                           & \SI{10}{\km} \\
    baryonic mass $M_b$                  & $\SI{1.25}{}M_\odot$ \\
    uniform baryonic density $\rho_b$    & \SI{3.3E+38}{\GeV\,\cm^{-3}} \\
    age $t_\text{NS}$                    & \SI{7}{\giga\year} \\
    escape velocity $\vesc(R)$ & \SI{1.8E+5}{\km \s^{-1}} \\
    number of neutrons $N_B$             & \SI{1.48E+57}{} \\
    temperature $T_\text{NS}$            & \SI{E+5}{\kelvin}
  \end{tabular}
  \end{ruledtabular}
\end{table}

%-----------------------------------------------------------------------------
\section{Constraints on repulsive forces}
\label{sec:repulsive}
%-----------------------------------------------------------------------------

\begin{figure*}
  \centering
  \includegraphics{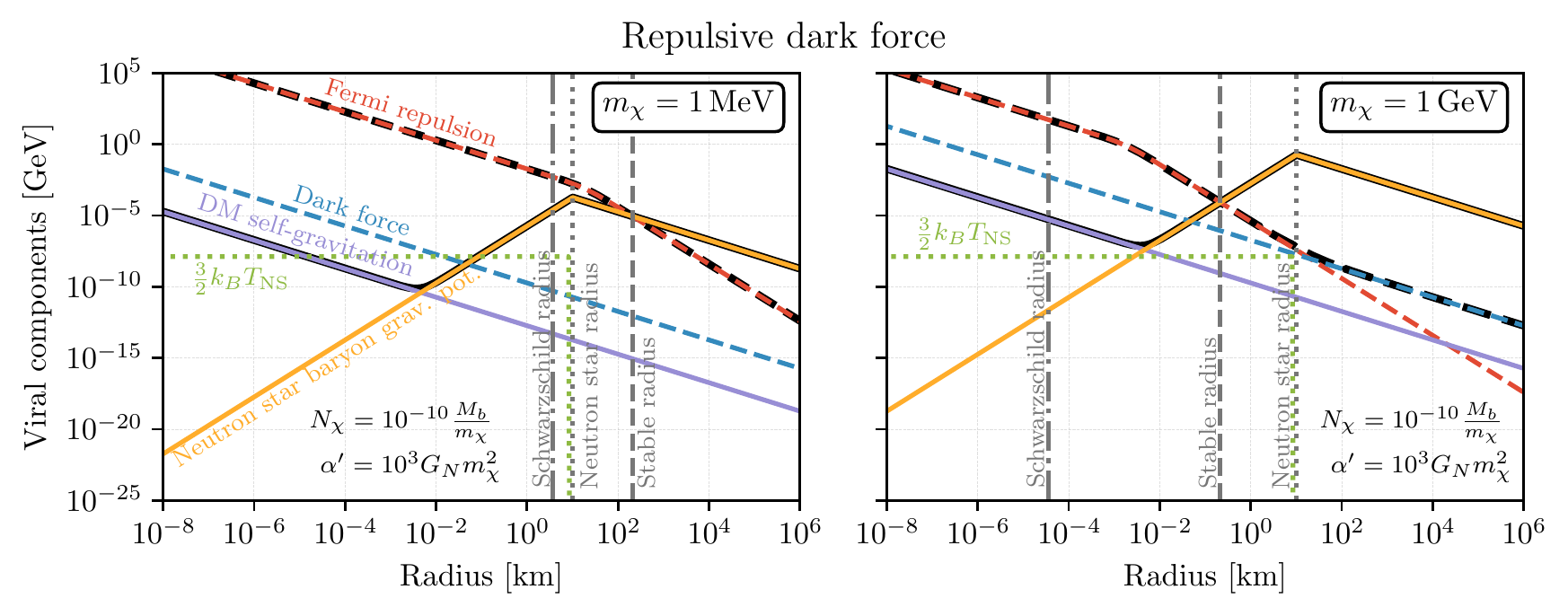}
  \caption{Visualization of the different contributions to the virial equation,
  \cref{eq:virial_eqn_substituted_repulsive}, as a function of the radius for 
  two different choices of the DM mass.  We plot the absolute values of the respective
  contributions, and indicate attractive (repulsive) forces as solid (dashed)
  lines. The black solid (black dashed) line correspond to the sum of all
  attractive (repulsive) forces. The solution to \cref{eq:virial_eqn_substituted_repulsive},
  i.e.\ the intersection of the black solid and black dashed lines, is
  highlighted by a vertical dashed line.}
  \label{fig:virialcomponents}
\end{figure*}

We now focus our attention on the case of a repulsive force between the dark
particles captured by binary neutron stars.  Our results on this scenario are
summarized in \cref{fig:Allbounds_repulsive_alpha_Nchi}.  The four panels of
this figure show, for different assumptions on the particle mass $m_\chi$, the
limits on the dark fine structure constant \alp and the total mass of all
$\chi$ particles in the neutron star, $M_\chi$.  We normalize \alp to the
effective strength of gravity, $G_N m_\chi^2$, and $M_\chi$ to the mass of the
baryons in the neutron star, $M_b$.  In the following, we discuss the various
constraints shown in \cref{fig:Allbounds_repulsive_alpha_Nchi} one by one.

Throughout our discussion of constraints we shall quote bounds on the dark
force parameter \altp under the assumption that both neutron stars have
identical dark charges and masses.  In the opposite limit that one neutron star
carries a dark charge while the other does not, equal constraints apply on the
parameter \btp.  It is crucial here that we have defined \btp such that under
the assumption that only one neutron star is charged its value is equal to that
of \altp for equally charged neutron stars.

\subsection{Binding potential for DM particles}
\label{sec:binding-potential}
%----------------------------------------------

It is intuitively clear that a repulsive force between dark particles will
limit the maximum number of such particles that can accumulate in or around a
neutron star or black hole.  Irrespective of the capture or production
mechanism, the number of particles is saturated once the net potential of the
compact object ceases to be attractive.  Consider the potential $V(r)$ felt by
a single DM particle $\chi$ of charge $q_\chi$ and mass $m_\chi$ at a distance $r$
from the center of the star:
\begin{align}
  V(r) &= - \frac{G_N \, M \, m_\chi}{r}
          + \frac{\alp \qchi Q \, e^{-\mmed r}}{r}\,.
\end{align}
Here, $Q$ is the total dark charge of the star, and
\begin{align}
  M = M_b + M_\chi
\end{align}
is its total mass (including the total baryonic mass, $M_b$, and the total mass
of dark particles, $M_\chi$). $M_\chi$ can be traded for the
number of particles, $N_\chi$, of a given mass $m_\chi$:
\begin{align}
  M_\chi = m_\chi N_\chi \,.
\end{align}
Insisting that the potential is attractive ($V(r) \leq 0$) at length scales
$r \ll \mmed^{-1}$ (where the Yukawa force is effectively Coulomb-like)
yields the condition
\begin{align}
  N_\chi &\lesssim \frac{M_b \, m_\chi}{\alp \, q^2_{\chi} / G_N - m_\chi^2} \,.
  \label{eq:maxNdm_replV}
\end{align}
This constraint is shown in light blue in \cref{fig:Allbounds_repulsive_alpha_Nchi}.

We can use \cref{eq:maxNdm_replV} to derive a constraint on the effective
strength \altp of the new force relative to gravity. Plugging
\cref{eq:maxNdm_replV} into \cref{eq:altp_definition}, we obtain
\begin{align}
  \altp \lesssim \frac{G_N m_\chi^2}{\alp \qchi ^2}\,.
  \label{eq:altp_repulsive_bound}
\end{align}

\subsection{Size of the Dark Core}
\label{sec:core-dynamics}
%---------------------------------

\begin{figure*}
  \centering
  \includegraphics[width=0.95\linewidth]{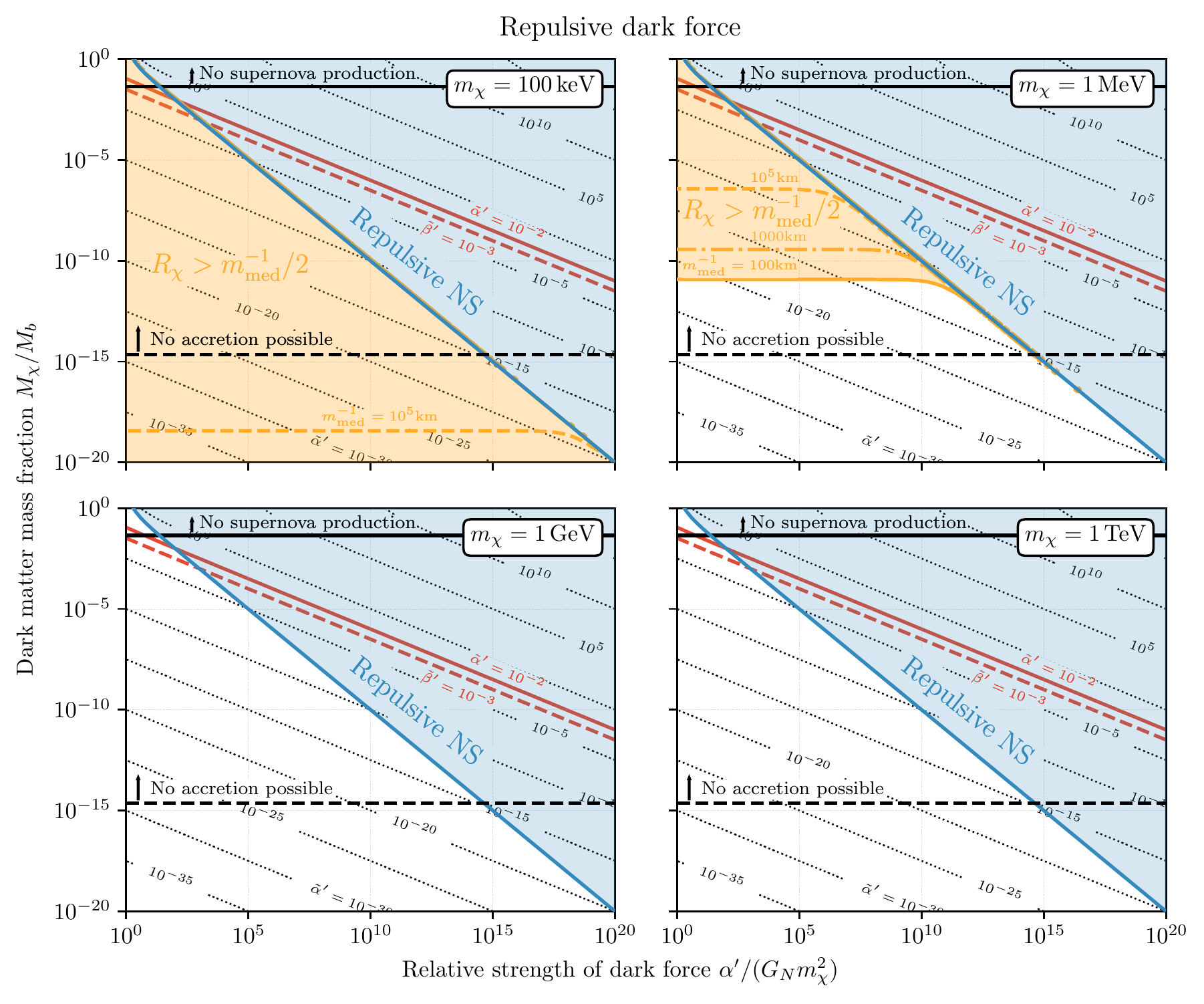}
  \caption{Constraints on the coupling \alp of a repulsive dark force
    as a function of the total mass $M_\chi$ of dark particles bound to the
    neutron star.  We normalize \alp to the strength of gravity, $G_N
    m_\chi^2$,  and $M_\chi$ to the baryonic mass of the neutron star, $M_b$.  The
    four panels correspond to different choices of $m_\chi$, as indicated in
    the plots. The dotted contours show the value of \altp (in the case of
    neutron stars with equal dark charge-over-mass ratio) or \btp (for the case that
    only one neutron star carries a dark charge, see \cref{eq:altp_definition})
    at each point in the plot. We indicate in red
    the estimated minimum value of \altp or \btp required for the dark force to leave
    an observable imprint in LIGO data.  Regions of parameter
    space where the neutron star no longer has an overall attractive potential
    for DM, i.e., $N_\chi$ exceeds the bound of \cref{eq:maxNdm_replV}, are
    shaded in blue. Note that this bound is independent of the DM production or
    capture mechanism. Bounds shown in black are model-dependent: in the region
    above the black-dashed line, the DM population inside the neutron cannot be
    generated by capture from the halo, based on the geometric arguments that
    lead to \cref{eq:maxNdm_dmcapt}.  Above the black solid line, the
    hypothesis that the DM population is produced during the
    supernova that created the neutron star is not tenable because the energy
    available in a supernova (as measured in SN~1987A) is not sufficient.
    Inside the yellow shaded region, the dark core has a radius larger than the
    range of new Yukawa force, $R_\chi \geq \mmed^{-1}/2$.  The solid, dot-dashed, and dashed
    yellow lines correspond to $\mmed^{-1} = \{\SI{100}{\km}, \SI{1000}{\km},
    \SI{1E+5}{\km}\}$, respectively. Within the yellow regions, the DM particles would
    be stripped away from their host neutron star before they can have a
    significant impact on the inspiral. The neutron star parameters are chosen in
    accordance with \cref{tab:neutron starparameters}, and we normalize \alp
    such that $\qchi = 1$. 
  }
  \label{fig:Allbounds_repulsive_alpha_Nchi}
\end{figure*}

In order for the effect of the dark force on the neutron star to be predictable
using the calculations of \cref{sec:inspiral_signal}, it is necessary that the
core of dark sector particles be significantly smaller than the screening radius of the
force.  Otherwise, the interactions of the neutron stars will be primarily with a
cloud of dark particles in which they are both embedded, rather than remaining
a simple central-force problem. This would require a much more sophisticated treatment.
We leave the dynamics of such a system for future work, and here aim to enforce
the requirement that the dark particle core be smaller in extent than the
screening radius $\mmed^{-1}$.

We remark however that the far-outlying particles at the edge of
the neutron stars' dark matter halo will be very easily stripped off in the
presence of the second neutron star because they feel the dark force field
created by that star's DM halo long before the inspiral dynamics can be
affected.  Thus, those outlying particles would not affect the gravitational
wave signal, as they would no longer be bound to just one of the two stars.

Using the virial theorem, we can calculate the stable core radius for given
\alp, $m_\chi$, and $N_\chi$. The general form of the virial theorem
(valid for non-relativistic and relativistic systems)
is $\ev{\,\sum_i \vec{p}_i \cdot \vec{v}_i} = -\ev{\,\sum_i \vec{F}_i \cdot
\vec{r}_i}$ \cite{goldstein1980classical},
where $i$ runs over all particles,
and $\vec{r}_i$, $\vec{p}_i$, $\vec{F}_i$ are their positions, their momenta, and
the forces they experience, respectively. The notation $\ev{\dots}$ indicates time
averaging, which here translates to the requirement that
the system is in a stable configuration. For an isotropic potential, the
virial equation in its relativistic form simplifies to
\cite{PhysRevLett.64.2733,rose_book}
\begin{align}
  \left\langle \sum_i \frac{\vec{p}_i^2}{\sqrt{\vec{p}_i^2 + m_\chi^2}} \right\rangle
    &= \left\langle \sum_i r_i \frac{\partial V_i(r_i)}{\partial r_i} \right\rangle \,.
  \label{eq:virialeqn}
\end{align}
The potential energy for a single particle at the edge of the dark core (radius $R_\chi$)
is given by 
\begin{align}
  \label{eq:rdeppot}
  V(R_\chi) &= -\frac{(G_N  m^2_\chi - \alp) N_{\chi}}{R_\chi}
                                                                \notag\\
            &\quad -G_N M_b m_\chi
                     \begin{cases}
                       (3 R^2 - R_\chi^2)/(2 R^3)
                                        & R_\chi \leq R \\
                       1/R_\chi
                                        & R_\chi > R
                     \end{cases} \,.
\end{align}
The first term is the combination of the gravitational self-interactions of the
exotic particles and the new dark interactions in the zero screening limit
$R_\chi\, \mmed \ll 1$. The second term is the gravitational interaction
between the baryons and the test particle defined separately in the two cases where the
$\chi$-core is either smaller or larger than the radius $R$ of the baryonic matter.
We assume that the dominant contribution to the average momentum on the left hand side
of \cref{eq:virialeqn} arises from Pauli repulsion. (We will justify this assumption
shortly.)
The $\vec{p}_i$ are then of the order of the Fermi momentum of a degenerate
3-dimensional system of free spin-$1/2$ fermions (with no additional internal
degrees of freedom)
\begin{align}
  p_F &= \left(\frac{3\pi^2 N_{\chi}}{V}\right)^{1/3}
       = \frac{1}{R_\chi} \left(\frac{9 \pi N_{\chi}}{4}\right)^{1/3}\,,
  \label{eq:fermienergy}
\end{align}
where $V$ is the volume of the $\chi$-core.  For repulsive forces, we find that the system is always in a
non-relativistic configuration in the parameter regions of interest to us
(unshaded portions of \cref{fig:Allbounds_repulsive_alpha_Nchi}),
allowing the simplification
\begin{align}
  \frac{p_F^2}{\sqrt{p_F^2 + m_\chi^2}}
    &= \frac{1}{m_\chi R_\chi^2} \left( \frac{9 \pi N_{\chi}}{4} \right)^{2/3}\,.
  \label{eq:fermienergy_simplified}
\end{align}
Inserting \cref{eq:fermienergy_simplified} and \cref{eq:rdeppot} into the
virial theorem, \cref{eq:virialeqn}, yields 
\begin{align}
  &\frac{1}{m_\chi R_\chi^2} \left( \frac{9 \pi N_\chi}{4} \right)^{2/3}
    = \frac{\left(G_N m^2_\chi - \alp\right) N_{\chi}}{R_\chi}  \notag\\
    &\qquad + G_N M_b m_\chi \begin{cases}
                                       R_\chi / R^3 & R_\chi \leq R \\
                                       1/R_\chi^2             & R_\chi > R
                                     \end{cases} \,.
  \label{eq:virial_eqn_substituted_repulsive}
\end{align}
The different contributions to this equation are represented graphically in
\cref{fig:virialcomponents}, where dashed lines indicate repulsive forces and
solid lines indicate attractive forces.  The plots also show the contribution
from ordinary thermal pressure, which, however, is relevant only
for relatively heavy DM ($m_\chi \gtrsim \text{TeV}$), weak dark forces
($\alp \sim G_N m_\chi^2$), and small dark cores ($M_\chi \lesssim 10^{-20} M_b$).
This parameter region is not of interest to us as it corresponds to tiny \altp; this justifies our previous assumption that degeneracy pressure dominates.
The intersection between the sum of all
repulsive contributions and the sum of all attractive contributions corresponds
to the solution of \cref{eq:virial_eqn_substituted_repulsive} and thus gives
the radius of the dark core as a function of $m_\chi$, $\alp$, and $N_\chi$.

For light DM mass, Fermi-repulsion results in radii larger than the size of
the neutron star. The regions of parameter space where $R_\chi > \mmed^{-1}/2$
are shaded in yellow in \cref{fig:Allbounds_repulsive_alpha_Nchi}.  A
$\chi$-core with such an extent results in inspiral dynamics very different to
those assumed in our estimates so far. In particular, as argued above,
particles at radii larger than $\mmed^{-1}$ will be efficiently stripped from
the neutron star.  The yellow curves, therefore, can be taken as an indication
of the maximum amount of charge that fits geometrically within the domain of
influence of a single neutron star, and thus can contribute to a dark-force-
induced change in inspiral dynamics and gravitational wave signals.

\subsection{Constraints on Dark Core Production via Particle Accretion}
\label{sec:accretion}
%----------------------------------------------------------------------

There exists an additional upper bound on $N_\chi$ under the assumption that
the particles giving the neutron star its dark charge have been accreted from the host
galaxy's DM halo.  An upper limit on the number $N_\chi^\text{capt}$ of DM particles
($\chi$)  accreted can be obtained by assuming that any DM particle
that passes through the neutron star over its
lifetime $t_\text{NS}$ is captured.  (A more realistic estimate
based on the dynamics
of baryon--$\chi$ scattering is discussed in \cref{appendix:DMcaptrate}.)
The upper limit on $N_\chi^\text{capt}$ is based on the
effective geometric cross section of the neutron star, $\sigma_\text{geom}^\text{eff}$,
given by
\begin{align}
  \sigma_\text{geom}^\text{eff}
    \equiv \sigma_\text{geom} \Big( 1 + \frac{\vesc^2}{\bar{v}^2} \Big)
    \simeq \frac{\sigma_\text{geom} \vesc^2}{\bar{v}^2}\,,
  \label{eq:sigma-geom-eff}
\end{align}
where $\vesc$ is the escape velocity at the surface of the neutron star and
$\bar{v}$ is the average $\chi$ velocity relative to the neutron star at
infinity. The velocity-dependent term by which $\sigma_\text{geom}^\text{eff}$
differs from the geometric cross section $\sigma_\text{geom} = \pi R^2$
accounts for the focusing of the DM wind by the neutron star's gravitational
field~\cite{Griest:1987vc}. In the second equality in
\cref{eq:sigma-geom-eff}, we use $v_\text{esc} > \bar{v}$, an approximation
that is well satisfied for typical galactic DM velocities $\bar{v} \sim
\SI{200}{\km\per\s}$ and escape velocities at the surface of a neutron star
$\vesc \sim \SI{100\,000}{\km\per\s}$.  The upper limit on $N_\chi^\text{capt}$
is then
\begin{align}
  N_\chi^\text{capt} &\leq \frac{\pi R^2 v^2_\text{esc} t_\text{NS} \rho_\chi}
                                {\bar{v} \, m_\chi} \,,
  \label{eq:maxNdm_dmcapt}
\end{align}
where $\rho_\chi$ is the mass density of $\chi$ in the vicinity of the neutron
star.  This constraint is shown as a horizontal black dashed line in
\cref{fig:Allbounds_repulsive_alpha_Nchi}.

As the constraints on $N_\chi$ from \cref{eq:maxNdm_replV,eq:maxNdm_dmcapt} scale
inversely to one another with respect to $m_\chi$, the value $m_\chi^\text{opt}$
that maximizes $N_\chi$ and thus \altp occurs when both constraints are equal.
We find
\begin{align}
  (m_\chi^\text{opt})^2 = \frac{\pi R^2 \vesc^2 t_\text{NS} \rho_\chi \alp q_\chi^2}
                               {G_N (M_b \bar{v}
                                   + \pi R^2 \vesc^2 t_\text{NS} \rho_\chi)}
\end{align}
and, by plugging $m_\chi^\text{opt}$ into \cref{eq:altp_repulsive_bound},
\begin{align}
  \altp &\leq \frac{\pi R^2 \vesc^2 t_\text{NS} \rho_\chi}
                   {M_b \bar{v} + \pi R^2 \vesc^2 t_\text{NS} \rho_\chi}
       \simeq \frac{\pi R^2 \vesc^2 t_\text{NS} \rho_\chi}{M_b \bar{v}}
                                                              \notag\\[0.2cm]
       &= \SI{2.25E-15}{} \,.
  \label{eq:repulsivebound}
\end{align}
The second equality follows because the mass of the accreted DM particles is
much smaller than the baryonic mass $M_b$.  For the numerical estimate in the second line
of \cref{eq:repulsivebound}, we have used the neutron star parameters given in
\cref{tab:neutron starparameters}, we have (very conservatively) assumed that
$\rho_\chi$ saturates the
observed DM density in the solar neighborhood $\rho_\chi =
\SI{0.3}{\GeV\per\cm^3}$, and we have taken $\bar{v} = \SI{220}{\km\per\s}$.
We note that this limit on \altp is independent of both $m_\chi$ and $\alp$.  
The limit is relaxed in regions of very low DM velocity $\bar{v}$.

\subsection{Constraints on Dark Core Production in Supernovae}
\label{sec:sn-repulsive}
%-----------------------------------------------------------------------

The upper limit on $N_\chi$ from \cref{eq:maxNdm_dmcapt}, and the resulting limit
on \altp, \cref{eq:repulsivebound}, were based on the assumption that the DM population
bound to the neutron star arises from capture of DM particles from the halo.
Relaxing this assumption and invoking more speculative production mechanisms can
lead to significant boosts in the allowed values of \altp.  We first note that
the largest \altp values compatible with the constraints of
\cref{sec:binding-potential} (blue regions in
\cref{fig:Allbounds_repulsive_alpha_Nchi}) occur in the region of parameter
space where the dark force is not much stronger than gravity and where $M_\chi
\sim M_b$.

One possibility for producing a massive dark core without accreting DM from the
halo arises if DM particles are produced
via bremsstrahlung in a new-born hot neutron star
\cite{Nelson:2018xtr,Ellis:2018bkr}. The
upper limit on the total mass of exotic particles produced this way arises from
the measured neutrino burst of SN~1987A \cite{Hirata:1987hu, Hirata:1988ad,
Bionta:1987qt, Bratton:1988ww, Alekseev:1988gp}.  Adopting the `Raffelt
criterion' \cite{Raffelt:1996wa} that $\chi$ production should not reduce
the duration of the neutrino burst from SN~1987A by more than 50\%, we require
that the instantaneous $\chi$ luminosity should be $L_\chi \leq \SI{3E+52}{\ergs\per\s}$.
Various different effects need to be taken into account to evaluate $L_\chi$
\cite{Rrapaj:2015wgs, Chang:2016ntp, Hardy:2016kme, Mahoney:2017jqk,
Chang:2018rso}. A near
future detection of supernova neutrinos in all flavors will robustly strengthen
these bounds~\cite{Laha:2013hva, Laha:2014yua, Nikrant:2017nya}.  

We
conservatively estimate that the total available energy to produce $\chi$
during the supernova is \SI{E+53}{\ergs}, which, under the further conservative
assumption that these particles are produced at rest, translates to $M_\chi
\lesssim \SI{0.045}{}M_\odot$. This constraint is very weak due to the current scarcity of supernova neutrino measurements, and corresponds to approximately 1/3 of the total energy of the supernova being used to produce DM. This constraint is shown as a solid black horizontal
line in \cref{fig:Allbounds_repulsive_alpha_Nchi}.  We can turn it into a constraint on
$\altp$ by requiring that it is saturated simultaneously with the the upper
bound on $N_\chi$ from \cref{eq:maxNdm_replV} (blue shaded region in
\cref{fig:Allbounds_repulsive_alpha_Nchi}).  The resulting equation can be
solved for \alp, and the result can be plugged into the definition of \altp,
\cref{eq:altp_definition}. If we assume that the two neutron stars have
similar mass and dark charge, we obtain the $m_\chi$-independent limit
\begin{align}
  \altp \leq \SI{4E-2}{} \,.
\end{align}

Note that the known thermal mechanisms to generate exotic particles in
supernovae are inefficient for particle masses $\gtrsim \SI{10}{MeV}$. In
\cref{fig:Allbounds_repulsive_alpha_Nchi}, we nevertheless plot the supernova
constraint for $m_\chi \gtrsim \SI{10}{MeV}$ as an absolute upper limit on the
amount of energy available to produce exotic particles in the supernova. We
are, however, not aware of any mechanism that could possibly be efficient in
converting an $\mathcal{O}\left( 30 \% \right)$ fraction of the supernova
energy into heavy particles. Any supernova-driven mechanism for DM production
would lead to all neutron stars being similarly charged, and thus the LIGO
sensitivity to these mechanisms is best modeled in
\cref{fig:Frequency_deviation}, as the charge-to-mass ratios of the binary
stars will be similar.

\subsection{Constraints on Dark Core Production via Neutron Decay}
\label{sec:n-decay}
%-----------------------------------------------------------------

An alternative production mechanism for DM particles inside a neutron star has
been proposed in Ref.~\cite{McKeen:2018xwc} and further investigated in
Ref.~\cite{Cline:2018ami}.  In this scenario, $\chi$ is abundantly produced
inside the neutron star via a hypothetical exotic neutron decay mode.  In this
case a repulsive dark force would have an additional benefit: without it, the
neutron star equation of state (EOS) would be softened because of decreased
Fermi repulsion in the nuclear matter.  This would preclude the existence of
high-mass neutron stars with $M \simeq 2 M_\odot$, in conflict with
observations \cite{Baym:2018ljz,Motta:2018rxp}.  A repulsive dark force,
however, stiffens the equation of state again.  This places a lower bound on
the size of the dark fine structure constant \alp. Ref.~\cite{Cline:2018ami}
found a bound on the strength of the needed repulsive force as a function of
the mediator mass.  In our notation, this constraint reads
\begin{align}
  \alp \geq \SI{1.5E-40}{} \left(\frac{\mmed}{\SI{100}{\km}}\right)^2\,.
  \label{eq:alp-constraint-Cline}
\end{align}
Note, however, that the calculations of Ref.~\cite{Cline:2018ami} were
performed in a scenario with a partially screened dark force, whereas scenarios
in which the dark force could significantly impact inspiral dynamics and
gravitational wave emission would require it to be unscreened inside
the neutron star.  This will weaken the constraint on \alp from EOS arguments
compared to \cref{eq:alp-constraint-Cline}.
Moreover, the neutron decay scenario requires careful tuning of
$m_\chi$: stability of ${}^9$Be places an upper bound on the mass splitting
between $\chi$ and the neutron of \SI{1.59}{\MeV}
\cite{McKeen:2015cuz,Fornal:2018eol}, and the lifetime of ${}^{11}$Be
further strengthens this bound to require splittings of less than
\SI{0.50}{\MeV} \cite{Ejiri:2018dun}.

To estimate the maximum quantity of hidden sector particles that can be produced
via neutron decay, we consider the chemical potentials of neutrons and DM.
We describe the neutrons by a Fermi gas with chemical potential
\begin{align}
  \mu_B = \sqrt{m_n^2 + (3\pi^2 n_n)^{2/3}} - E_B \,,
  \label{eq:n-decay-mu-B}
\end{align}
where $n_n$ is the number density of neutrons (which decreases over time in this
scenario), and $E_B$ is a typical nuclear binding
energy, taken here to be $E_B = \SI{9}{\MeV}$ \cite{Cline:2018ami}.
The second term under the square root corresponds to the square of the neutron
Fermi momentum.

We assume the $\chi$ particles experience an unscreened potential, i.e.,
$\mmed^{-1} \gg R$, resulting in the chemical potential
\begin{align}
  \mu_\chi &\simeq m_\chi + \frac{\alp N_\chi}{R_\chi} \,.
\end{align}
Neutron decays occur until $\mu_\chi = \mu_B$. Therefore, the maximum $N_\chi$
can be determined by equating the two chemical potentials.  Choosing for
$m_\chi$ the smallest value allowed by the lifetime of \iso{Be}{11}, $m_\chi =
\SI{939.06}{\MeV}$ \cite{Ejiri:2018dun}, and for \alp the smallest value which
can still be comparable in its effects to those of gravity, $\alp = G_N
m_\chi^2$, we obtain $M_\chi / M_b \leq 0.36$. Here, we have used the neutron
star parameters given in \cref{tab:neutron starparameters}, and we have assumed the
radius of the dark core to be similar to the radius of the neutron star. 

The above constraint on $M_\chi / M_b$ corresponds to $\altp = 0.071$, which
would be marginally detectable in LIGO/VIRGO data, see
\cref{fig:Frequency_deviation}. Such a modification could be more clearly
observable by comparing LIGO/VIRGO measurements with results from other lower
frequency instruments\,\cite{Alexander:2018qzg}. Note that the upper bound on
\altp in the neutron decay scenario has been calculated for the choices of
$m_\chi$, $\alp$ which maximize the potential effect on inspirals; either
increasing $m_\chi$ or increasing $\alp$ leads to a tighter upper bound on the
mass ratio $M_\chi / M_b$. An increased $\alp$ thus leads to a decrease in the
dark charge, so that overall the strength of the dark force between the binary
stars is decreased. Note also that, in scenarios where dark particles are
produced mainly via exotic neutron decay, there is no way to create an
uncharged neutron star, so the dominant effect of the dark force is always a
modification of the potential in which the neutron stars orbit as shown in
\cref{fig:Frequency_deviation}, never dipole radiation of the force mediator.

We finally note that the coupling that induces decays of the form $n \to \chi + X$
(where $X$ can, but does not have to be, the dark force mediator $A'$) will
always lead to a coupling of neutrons to $A'$ at 1-loop level. This shows that
the assumption of a dark force coupling only to hidden sector particles must be
relaxed to produce an observable effect in the inspiral waveform.  Then,
however, new physics constraints from modified gravitational wave signals need
to be discussed in the context of constraints arising from more generic
fifth-force searches \cite{Salumbides:2013dua, Salumbides:2013aga,
Adelberger:2009zz}.  The interplay between these various constraints is
model-dependent and would require dedicated studies.

%-----------------------------------------------------------------------------
\section{Constraints on attractive forces}
\label{sec:attractive}
%-----------------------------------------------------------------------------

We now switch gears and consider attractive dark sector
forces.  In this case, important constraints arise from black hole formation
inside neutron stars.  Additionally, the gravitational binding of DM particles
to the neutron star is now small compared to their attractive
self-interactions, leading to the possibility that the dark core can migrate
out of the neutron star and stop affecting the observed inspiral dynamics. In the
following we provide analytical estimates of these constraints and then present
the full numerical results in
\cref{fig:Allbounds_alpha_Nchi,fig:Allbounds_mchi_alpha}.

\subsection{Black hole formation}
\label{sec:bh-formation}
%--------------------------------

As a neutron star accumulates a large number of DM particles close to its
center, it runs the risk of the dense dark core collapsing into a black hole
which eventually consumes the whole star \cite{Bramante:2013nma,
Kouvaris:2011gb}.  The mere existence of old neutron stars tells us that black
hole formation cannot be too efficient, and this in turn restricts the DM
parameter space~\cite{2004hpa..book.....L, Manchester:2004bp}.  The most
optimistic scenario (i.e.\ the one in which black hole formation is least
likely) assumes that the DM particle $\chi$ is a fermion. In this case, Pauli
repulsion helps to stabilize the dark core at radii larger than the
Schwarzschild radius.

\begin{figure*}
  \centering
  \includegraphics[width=0.95\linewidth]{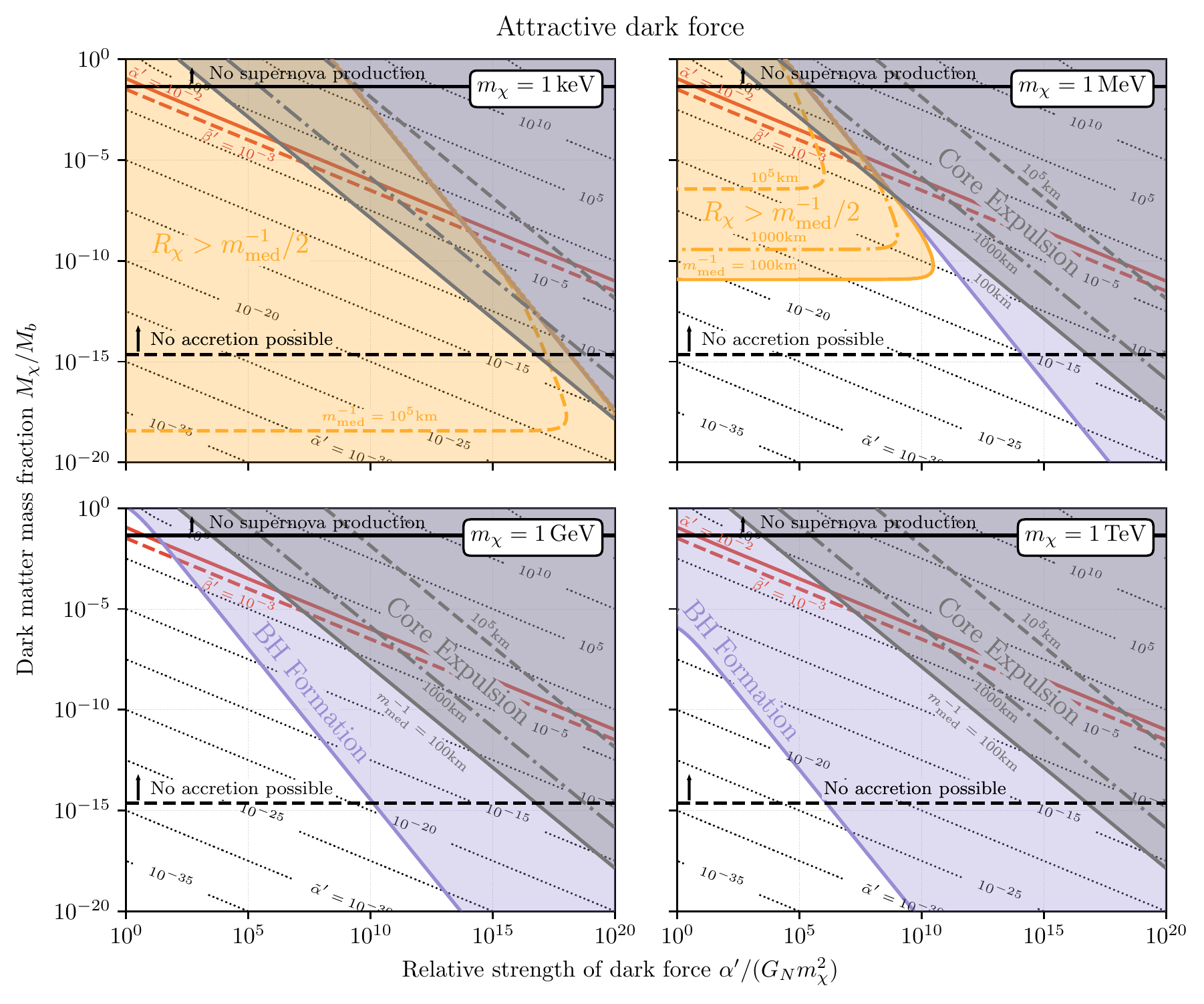}
  \caption{Constraints on an attractive dark force from binary neutron star systems
    as a function of the dark fine structure constant \alp (expressed here
    relative to the gravitational force) and the amount of DM bound to the
    neutron stars (assuming both stars carry the same amount of DM).  The four
    panels correspond to different choices of the DM mass $m_\chi$, as
    indicated in the plots. The dotted contours show the value of \altp (in the case of
    neutron stars with equal dark charge-over-mass ratio) or \btp (for the case that
    only one neutron star carries a dark charge, see \cref{eq:altp_definition})
    at each point in the plot. We indicate in red
    the estimated minimum value of \altp or \btp required for the dark force to leave
    an observable imprint in LIGO data.  In the purple
    region, the dark force induces collapse of the neutron stars' dark cores to
    black holes.  Inside the gray shaded region, the dark cores are expelled from the
    neutron stars as soon as the screening of the force is lifted.  We show this region for
    $\mmed^{-1} = \SI{100}{\km}$ (solid), $\SI{1000}{\km}$ (dot-dashed), and
    $\SI{1e5}{\km}$. Bounds shown in
    black are model-dependent: in the region above the black-dashed line,
    the DM population inside the neutron cannot be generated by capture from
    the halo, based on the geometric arguments that lead to
    \cref{eq:maxNdm_dmcapt}.  Above the solid black line, the hypothesis that
    the DM population is produced radiatively during the supernova that created
    the neutron star is not tenable because the energy available in a supernova
    (as measured in SN~1987A) is not sufficient.  Within the yellow regions, the DM particles would be stripped away from their host neutron star before they can have a significant impact on the inspiral.  The benchmark values of $\mmed^{-1}$ are the same as for the
    gray DM expulsion bound.  The neutron star parameters are chosen in
    accordance with \cref{tab:neutron starparameters}, and without loss of generality we
    have chosen $\qchi = 1$.}
  \label{fig:Allbounds_alpha_Nchi}
\end{figure*}

\begin{figure*}
  \centering
  \includegraphics{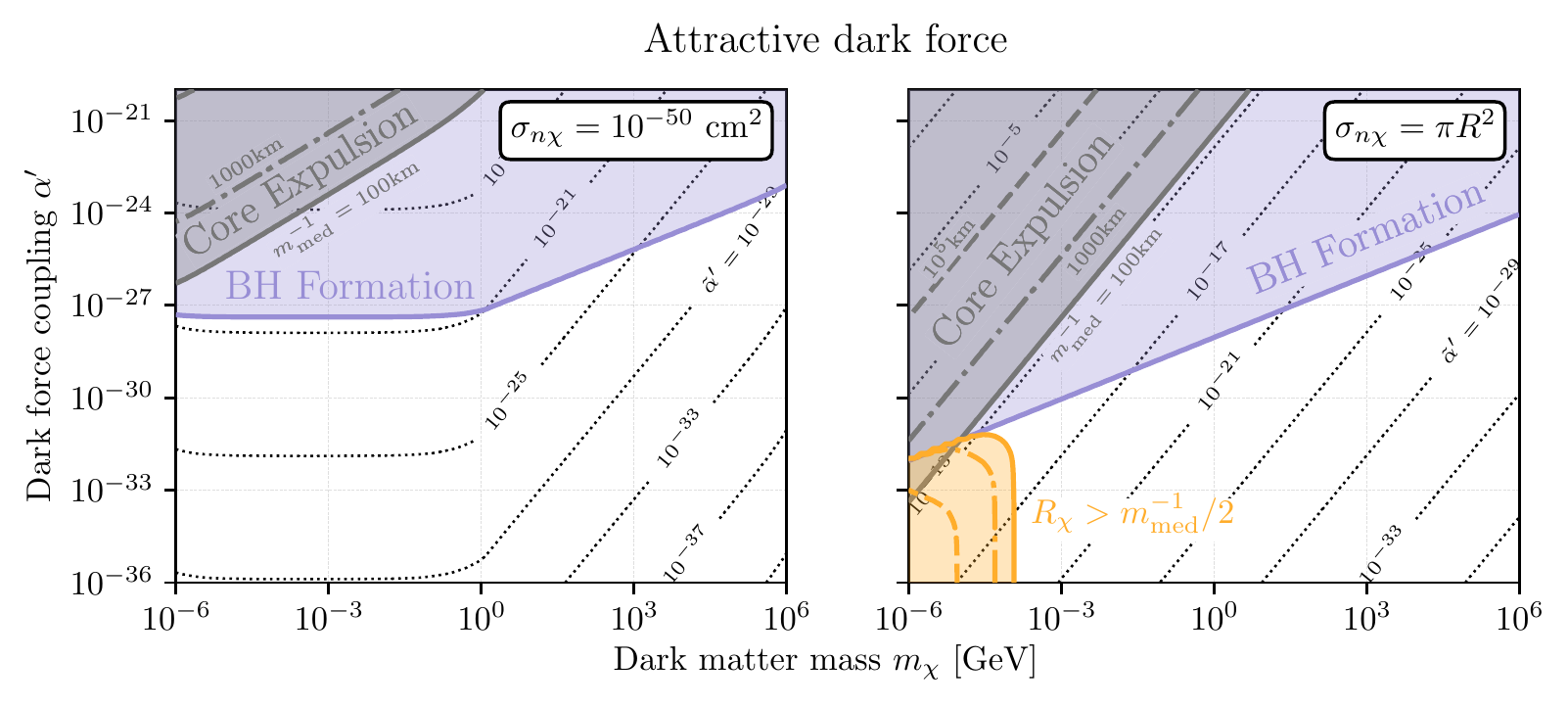}
  \caption{Constraints on an attractive dark force from a binary neutron star system
    as a function of the DM mass $m_\chi$ and the dark fine structure constant
    \alp.  The dotted contours show the value of \altp (in the case of
    neutron stars with equal dark charge-over-mass ratio) or \btp (for the case that
    only one neutron star carries a dark charge, see \cref{eq:altp_definition})
    at each point in the plot. In the purple
    region, the dark force induces collapse of the neutron stars' dark cores to
    black holes, assuming the amount of DM inside the stars is determined by DM
    capture from the halo, see \cref{appendix:DMcaptrate}.  Inside the gray shaded region, the dark cores are expelled from the
    neutron stars as soon as the screening of the force is lifted. 
    We show this region for $\mmed^{-1} = \SI{100}{\km}$ (solid),
    $\SI{1000}{\km}$ (dot-dashed), and $\SI{1e5}{\km}$.  Within the yellow regions, where $R_\chi \geq \mmed^{-1}/2$, the DM particles would be stripped away from their host neutron star before they can have a significant impact on the inspiral. The benchmark values of $\mmed^{-1}$ are the same as for the DM expulsion bound.  The neutron star parameters are chosen in accordance
    with \cref{tab:neutron starparameters}, and without loss of generality we have chosen
    $\qchi = 1$.  We have assumed a DM--nucleon scattering cross section
    $\sigma_{n\chi} = \SI{8E-47}{\cm^2}$, a DM density in the vicinity of the
    neutron star of $\rho_\chi = \SI{0.3}{\GeV\per\cm^3}$, and a DM velocity
    relative to the neutron star of $\bar{v} = \SI{220}{\km\per\s}$.}
  \label{fig:Allbounds_mchi_alpha}
\end{figure*}

For the following estimate we follow Refs.~\cite{McDermott:2011jp,
Bramante:2013nma, Kouvaris:2011gb}, but assuming that the range of the dark
force is always larger than the radius $R_\chi$ of the $\chi$ core.  Using the virial
theorem (see \cref{sec:core-dynamics}), we determine if a stable radius exists
for given values of \alp, $m_\chi$, and $N_\chi$.  In contrast to the case of
repulsive forces, it is possible for the DM particles to be either
non-relativistic, fully relativistic, or in between.  Inserting the Fermi
momentum $p_F = (9\pi N_\chi / 4)^{1/3} / R_\chi$ from \cref{eq:fermienergy},
and the potential energy given by \cref{eq:rdeppot} with the replacement $\alp
\to -\alp$, into the virial equation \cref{eq:virialeqn}, we obtain
\begin{multline}
  \frac{1}{R_\chi} \frac{\big( \tfrac{9}{4} \pi N_\chi \big)^{2/3}}
                        {\sqrt{\big( \tfrac{9}{4} \pi N_\chi \big)^{2/3} + m_\chi^2 R_\chi^2}}
    = \frac{\left(G_N m^2_\chi + \alp\right) N_{\chi}}{R_\chi}  \\
            + G_N M_b m_\chi \begin{cases}
                                       R_\chi / R^3 & R_\chi \leq R \\
                                       1/R_\chi^2             & R_\chi > R
                                     \end{cases} \,.
  \label{eq:virial_eqn_substituted}
\end{multline}
This expression is of course fully analogous to
\cref{eq:virial_eqn_substituted_repulsive}, and the different terms in it can
be visualized in \cref{fig:virialcomponents}, except that now the dark force is
attractive, i.e.\ the cyan lines should now be solid rather than dashed. If,
for a given $N_\chi$, a solution for $R_\chi$ exists, the dark core is stable.
If no solution exists, the core will collapse into a black hole. We have
checked that in the parameter regions of interest to us, the solution for
$R_\chi$, if it exists, is always larger than the Schwarzschild radius.

For the parameter ranges that we are most interested in, namely those featuring
the largest possible $N_\chi$ and \alp, minimum stable $R_\chi$ values
typically occur at relativistic $p_F$.  In this case, 
\cref{eq:virial_eqn_substituted} can be simplified to the constraint
\begin{align}
  N_\chi &< \frac{3\sqrt{\pi}}{2(\alp + G_N \, m_\chi^2)^{3/2}} \,.
  \label{eq:Nchi-constraint-BH-formation}
\end{align}
This limit is saturated if the baryonic gravitational potential is completely
negligible.
If we assume that the maximum value of $N_\chi$ is determined by DM capture
from the halo (see \cref{eq:maxNdm_dmcapt}), we can turn
\cref{eq:Nchi-constraint-BH-formation} into a constraint on \altp. Using
the definition of \altp from \cref{eq:altp_definition}, we obtain
\begin{align}
  \altp &\lesssim \left(\frac{9 \pi^2}{4} \right)^{1/3}
                  \frac{1}{G_N M^2}
                  \left( \frac{\pi R^2 \vesc^2 t_\text{NS} \rho_\chi}
                              {\bar{v} \, m_\chi} \right)^{4/3} \notag\\
        &= \SI{6.7E-20}{} \left(\frac{\SI{1}{\GeV}}{m_\chi}\right)^{4/3} \,.
  \label{eq:final_altp_limit}
\end{align}
Here, we have used the neutron star parameters from \cref{tab:neutron starparameters} for
both neutron stars.

The parameter values where no stable dark cores can exist, i.e.\ where
\cref{eq:virial_eqn_substituted} has no solution, are shown in purple in
\cref{fig:Allbounds_alpha_Nchi,fig:Allbounds_mchi_alpha}.  In
\cref{fig:Allbounds_alpha_Nchi}, no assumptions are made on how the dark
particles came to be within the neutron star.  In
\cref{fig:Allbounds_mchi_alpha}, the same constraints are shown as a function
of $m_\chi$ and $\alp$, assuming that the dark core has been accumulated via
capture.  The two panels correspond to different assumptions on the capture
cross section $\sigma_{n\chi}$.  We observe from
\cref{fig:Allbounds_mchi_alpha} that at $m_\chi \gtrsim \SI{1}{GeV}$, larger
$m_\chi$ implies more stringent constraints on \alp.  This is because of the lower DM
number density at large $m_\chi$, which implies smaller total dark charge
$Q_\chi$ and thus a smaller attractive force. Of course, Fermi repulsion is
also weaker when less DM is accreted, however the corresponding contribution to
the virial equation \eqref{eq:virial_eqn_substituted} scales less strongly with
$m_\chi$ than the dark force potential.  At $m_\chi \lesssim \SI{1}{GeV}$, the
purple exclusion curve and the $\altp$ contours in the left panel of
\cref{fig:Allbounds_mchi_alpha} flatten because of Pauli blocking, which limits
the number of final states available to the neutrons participating in
scattering processes and thus limits the number of target particles that DM
particles can scatter off and be captured.

The bounds from black hole formation shown in
\cref{fig:Allbounds_alpha_Nchi,fig:Allbounds_mchi_alpha} have been derived
assuming the dark core is a degenerate Fermi gas, i.e.\ that Fermi repulsion is
maximal.  It is possible that the dark core collapses to form a black hole before the
required number of particles are captured to satisfy the degeneracy condition.
This would lead to a more stringent bound in comparison to
\cref{fig:Allbounds_alpha_Nchi,fig:Allbounds_mchi_alpha}. We have also neglected
the question of whether the black hole formed would indeed consume the entire
neutron star in finite time or if it evaporates.
If the black hole evaporates
before consuming the star, no residual dark core would be left, so in this
case no modifications to the inspiral dynamics are expected.
However, using the results
of ref.~\cite{Capela:2013yf}, we conclude that evaporation is faster than
the collapse of the neutron star only if $M_\chi \lesssim \SI{1E-20}{\msun}$.

Solving \cref{eq:virial_eqn_substituted} also yields the radius of the dark
core when such a stable radius exists. For light DM mass, Fermi repulsion
results in core radii $R_\chi$ larger than the neutron star radius $R$.  For
very light DM, $R_\chi$ becomes even larger than $\mmed^{-1}$, so that the
dynamics of the dark core and the binary inspiral would be very different to
that of \cref{sec:inspiral_signal}. The parameter region where this happens is
shaded in yellow in \cref{fig:Allbounds_alpha_Nchi,fig:Allbounds_mchi_alpha}.
As previously discussed in \cref{sec:core-dynamics}, these yellow curves
correspond to the maximum amount of charge that fits within the domain of
influence of the neutron star. Particles at radii larger than $\mmed^{-1}$
will be efficiently stripped off the neutron star.

\subsection{Expulsion of the Dark Core}
%--------------------------------------

We can estimate whether the dark cores of two inspiraling neutron stars remain
trapped at the center of their host stars, or whether they are ejected from
them by virtue of the strong dark sector force acting on them.  If the dark
cores are ejected as soon as the dark force between the two neutron stars
becomes unscreened the observed signal will be unchanged compared to the
predictions of pure general relativity.

Consider the case where the dark core of one of the neutron stars
is about to be expelled from the star and is located
at its boundary.  We compare the gravitational force
experienced by the DM particles (which is strongest at the boundary of the neutron star)
to the opposing dark force induced by the dark core of the other
neutron star in the binary system:
\begin{align}
  |F_{\text{grav},i}| &= \frac{G_N M_i M_{\chi,i}}{R_i^2} \,, \\ 
  |F_{\text{dark},i}| &= \frac{\alp Q_i Q_j}{\Delta^2} (1 + \mmed\Delta) e^{-\mmed \Delta} \,.
\end{align}
Here the indices $i$ and $j$ label the two neutron stars ($i=1$, $j=2$ or
$i=2$, $j=1$).  As before, $R_i$ are the neutron star radii, $\Delta$ is the
distance between them, $M_{b,i}$ and $M_{\chi,i}$ are the total masses of their
baryonic and dark constituents, respectively, $M_i = M_{b,i} +
M_{\chi,i}$ are their total masses, and $Q_i$ are their dark charges.  The dark
core is expelled from the neutron star if $|F_{\text{grav},i}| <
|F_{\text{dark},i}|$, i.e.\ if
\begin{align}
  \left(\frac{\Delta}{R_i}\right)^2
    &< \frac{\alp Q_i Q_j (1 + \mmed\Delta) e^{-\mmed \Delta}}
            {G_N M_i M_{\chi,i}} \,.
  \label{eq:four-body_bound}
\end{align}
In the regime $\mmed\Delta \ll 1$, and for DM particles carrying unit charge
under the dark force ($q_\chi = 1$), this condition simplifies to
\begin{align}
  \left(\frac{\Delta}{R_i}\right)^2
    &< \frac{\alp N_{\chi,j}}
            {G_N M_i m_\chi}
    &\simeq 10^{25} \sqrt{\alp \altp} \left(\frac{\SI{1}{\keV}}{m_\chi}\right) \,.
\end{align}
In the last equality, we have assumed that both neutron stars carry equal
amounts of DM, and we have used the definition of \altp from \cref{eq:altp_definition}.
This implies that the gravitational potential can be overcome already at very
large distances, leading to expulsion of the dark cores long before the actual
merger.

In \cref{fig:Allbounds_alpha_Nchi,fig:Allbounds_mchi_alpha} we show the bounds
arising from requiring that the expulsion distance be less than the force range
as gray contours.  These contours are calculated without neglecting Yukawa
screening effects.  We consider three different values for the range of the
dark force, $\mmed^{-1}$: at $\mmed^{-1} = \SI{100}{km}$, the onset of the dark
force would happen within the LIGO/VIRGO frequency band; at $\mmed^{-1} =
\SI{1000}{km}$, the dark force would be suppressed while the binary system
emits gravitational waves in the lower frequency band, but unsuppressed when it
emits in the LIGO/VIRGO band; at $\mmed^{-1} = \SI{1e5}{km}$, the dark force
would be important for both low frequency detectors and LIGO/VIRGO.

\subsection{Constraints on Dark Core Production Mechanisms}
\label{sec:sn1987a}
%----------------------------------------------------------

As in the case of a repulsive dark force (\cref{sec:repulsive}), further
constraints exist on specific dark core production mechanisms.  In
\cref{fig:Allbounds_alpha_Nchi}, we indicate by the horizontal black dashed
lines the maximum amount of dark matter that can be accreted from the halo
assuming a geometric capture cross section (see \cref{sec:accretion} and
\cref{appendix:DMcaptrate}). \Cref{fig:Allbounds_mchi_alpha} also assumes DM
particle capture, with either a fixed DM--nucleon scattering cross section (left
panel) or a geometric capture cross section (right panel).
The greatest possible value of \altp that can be realized assuming the
dark core is produced via accretion from the halo (\cref{eq:maxNdm_dmcapt}) is
$\altp\sim \SI{E-14}{}$. It occurs for $m_\chi = \SI{0.1}{\MeV}$.  These values are
determined by calculating the $m_\chi$ value where the flat yellow region of
\cref{fig:Allbounds_alpha_Nchi} hits the accretion line and then plugging this
value into the black hole formation bound.

Producing dark sector particles in neutron decay is not a viable option for
attractive dark forces. In the absence of such a dark force the neutron star
equation of state is already too soft to allow for $2M_\odot$ neutron stars
\cite{McKeen:2018xwc,Cline:2018ami}, and an attractive force will only further
soften it.

The possibility that the dark core is produced radiatively during the supernova
explosion that creates the neutron star is constrained in the same way as for
repulsive dark forces, see \cref{sec:sn-repulsive}.  This constraint is shown in
\cref{fig:Allbounds_alpha_Nchi} as a solid horizontal black line.  The greatest
possible value of \altp for radiatively produced DM is \SI{0.23}{}. It occurs
for $m_\chi = \SI{0.25}{\GeV}$ under the assumption of perfectly efficient
production in the supernova.
Note again that these
masses are well above the supernova temperature, thus it is implausible
to expect a large fraction of the supernova energy to be converted to dark
particles in order to achieve this upper-limit value.

In summary, we conclude from \cref{fig:Allbounds_alpha_Nchi} that dark sector
forces cannot significantly affect gravitational wave observations of neutron
star inspirals as long as we assume that the neutron stars acquire their dark
charge during or after their creation in a supernova.
In principle, this no-go theorem could be avoided in tiny regions
of parameter space (upper left corner of the plots in \cref{fig:Allbounds_alpha_Nchi})
by a mechanism that endows neutron stars with dark cores that account for
at least $\sim 1\%$ of their mass.
Such a massive DM core could be realized if the DM is already collapsed into a very
dense structure, and subsequently the entire DM core either seeds the formation
of a star or is captured en masse by a star or neutron star in what must be an incredibly
rare event. Of course, such a mechanism is unavailable to repulsively-interacting DM,
as the dark sector interactions would prevent such collapse.

Both of these scenarios are constrained by the requirements that the collapse
of pure DM systems not damage the observed overall halo structure of galaxies.
In particular, collapsing and asymmetrically charged DM objects should only
account for a small fraction of the DM in the Universe, making the envisioned
capture events even more rare. A detailed analysis of the abundance bounds in
this case is left to a future publication. Assuming nonetheless that the
neutron star consists of equal parts DM and baryons, i.e.\ $M_\chi\sim M_b$,
the maximum value of $\altp\sim0.3$ occurs for $m_\chi \sim \SI{1}{\GeV}$ and
$\alp \sim G_N m_\chi^2$, with the amount of DM particles just below that
which would cause collapse to a black hole.

Since compact dark objects that  have collapsed in the early Universe are
scarce, it is likely in the above scenarios that a neutron stars carrying
a large dark charge forms a binary system with an uncharged partner.
Therefore, in these scenarios, it is likely that the dominant signature of
the dark force would be due to dipole radiation (see \cref{sec:radiation}).

%-----------------------------------------------------------------------------
\section{Conclusions}
\label{sec:conclusions}
%-----------------------------------------------------------------------------

We have investigated the conditions under which a new long-range force acting
on dark matter can affect the dynamics of neutron star inspirals and can lead
to observable modifications to gravitational wave signals observed in
gravitational wave detectors such as LIGO/VIRGO.  This scenario is based on the
assumption that a large population of DM particles exists within neutron stars.

For repulsive forces, a crucial constraint arises from the requirement that the
dark core of the neutron star remains stable (i.e.\ that gravitational
attraction remains stronger than the repulsive dark force).  Moreover, such
scenarios are severely limited by the fact that a sufficiently large dark core
cannot be produced via DM accretion from the halo; the greatest effect on
inspirals which is possible through accretion has strength normalized to gravity of
$\altp=\SI{2E-15}{}$. Considering more exotic production mechanisms for the DM
particles, we conclude that the strongest possible signal in the case of a
repulsive dark force corresponds to an \altp value of 0.071, and arises through
neutron decays similar to those which might explain the neutron lifetime
measurement disagreements.

For attractive dark forces additional constraints arise from the possible
collapse of a massive dark core to a black hole that would consume the whole
neutron star.  In addition, a strong attractive force could lead to the expulsion of
the dark cores from their host neutron stars in a binary system.  In this case,
the dark cores and their self-interaction cannot contribute significantly to
gravitational wave signals any more. A similar effect is the stripping of DM
particles from the neutron stars in a binary system. In fact, if the DM
population bound to a neutron star consists of relatively light particles, it
may extend to radii much larger than that of the baryonic matter.  Such DM
particles are easily freed from the star's binding force and thus stop
contributing to the inspiral dynamics.

In the case of attractive dark forces, the
maximum effect on the inspiral dynamics and the gravitational wave signals also
depends on the mechanisms by which the dark cores of the neutron stars form.
For accretion from the halo, the maximum possible effect is characterized by
$\altp = 10^{-14}$.  If the DM population is produced in the supernova that
also gives birth to the neutron star, the strongest possible signal is given by
$\altp = \SI{0.04}{}$.  Even stronger modifications to gravitational wave
signals are possible up to $\altp=0.3$. This value, however, can only be
realized if the DM particles have either seeded the formation of the neutron
star's progenitor star or have been accreted as an already collapsed core in a
very rare process.

We conclude that, if any deviations from the predictions of general relativity
are found by LIGO/VIRGO in gravitational wave signals from neutron star
inspirals, new exotic mechanisms of DM production in neutron stars are
required. This would likely herald the existence of large, compact DM
structures that either seed star formation or are captured by stars or neutron
stars later in their lives. It has been shown in Ref.~\cite{Alexander:2018qzg}
that future gravitational wave telescopes like the Einstein Telescope can
substantially improve the new physics sensitivity following a similar search
strategy.

We finally note that, beyond the observables discussed in this paper, the
dynamics of the dark sector could also affect gravitational waves from binary
neutron star inspirals in very different ways.  Namely, rather than altering
the force between the two inspiraling bodies, DM particles could instead alter
the properties of the neutron stars. This includes, for instance, alterations
in their tidal deformability due to the presence of dark sector particles
\cite{Nelson:2018xtr} or alterations arising from particles produced by black
hole superradiance effects \cite{Baumann:2018vus}. Neither of these effects
would contradict the results presented here for probing dark sector forces.

%-----------------------------------------------------------------------------
\section*{Acknowledgments}
%-----------------------------------------------------------------------------

The authors gratefully acknowledge helpful discussions with Chen Sun and Jun
Zhang.  We especially want to thank Ian Harry for many useful comments
on the manuscript.
The work of WS was supported by the Alexander von Humboldt Foundation in the
framework of the Sofja Kovalevskaja Award 2016, endowed by the German Federal
Ministry of Education and Research. JK, RL, and TO have received funding from
the German Research Foundation (DFG) under Grant Nos.\ EXC-1098, FOR~2239 and
GRK~1581, and from the European Research Council (ERC) under the European
Union's Horizon 2020 research and innovation programme (grant agreement No.\
637506, ``$\nu$Directions''). JK and TO would like to thank the CERN Theoretical
Physics Department for hospitality and support.

%-----------------------------------------------------------------------------
\appendix 
%-----------------------------------------------------------------------------
\section{Capture rate calculation}
\label{appendix:DMcaptrate}
%-----------------------------------------------------------------------------

The capture rates for dark particles in the scenarios under consideration in
this paper have been calculated in refs.~\cite{Gould:1987ju, Gould:1987ir,
McDermott:2011jp, Bramante:2013nma}.
Here, we summarize these calculations, fixing the notation and highlighting
the various assumptions made. We emphasize that, while we have utilized standard
DM halo parameters throughout this article as a benchmark, our results do not
require that the particles captured in neutron stars make up all the DM in the
Universe.

Firstly, we approximate that the average DM velocity, escape velocity, and
baryon density are uniform throughout the body of the neutron star and that the
DM--nucleon scattering cross section is velocity-independent. Then, the capture
rate of DM on a neutron star is
\begin{align}
  C_\chi &\simeq \sqrt{\frac{6}{\pi}} \frac{\rho_\chi}{m_\chi}\frac{1}{\bar{v}}\frac{v(r)^2}{1-v(r)^2}
                 f(\sigma_{n\chi}) \xi N_B \bigg[ 1-\frac{1-e^{-B^2}}{B^2} \bigg]\,.
  \label{eq:C-chi}
\end{align}
The parameters and functions that enter this equation and have not already
been defined are:
\begin{itemize}
  \item $v(r)$ is the infall speed of the dark particles, which we approximate by the
    escape velocity at the surface of the neutron star.  For compact objects
    like neutron stars, this is a good approximation because the kinetic energy
    a dark particle acquires during infall is much larger than the initial kinetic
    energy at $r = \infty$.
    
    \item The factor $1-v(r)^2$ in the denominator arises from
general relativity corrections in the vicinity of the neutron star
\cite{Kouvaris:2007ay}. These leads to a $\sim60\%$ increase of the capture
rate for the neutron star parameters in  \cref{tab:neutron starparameters}.

  \item $\xi$ is the fraction of neutrons that contribute in the scatting after
    including the effects of Pauli blocking. $\xi=1$ for DM masses above a GeV,
    but for smaller values where the momentum of the DM particle becomes smaller
    than the neutron Fermi momentum  $p_F = (3\pi^2 \rho_n/m_n)^{1/3}$ (given
    here in terms of the neutron mass density $\rho_n$ and the neutron
    mass $m_n$) the fraction of neutrons which have enough momentum to have
    an accessible final state to scatter in to is
    \begin{align}
      \xi \simeq \frac{\sqrt{2} m_r v_\text{esc}}{p_F}\,. 
    \end{align}
    Here $m_r$ is the reduced mass of the neutron--DM system, $m_r = m_\chi
    m_n/(m_\chi + m_n)$.

  \item $f(\sigma_{n\chi})$ is a function of the DM--nucleon cross section that
    determines the probability of scattering:
    \begin{align}
      f(\sigma_{n\chi}) &= \sigma_\text{sat}
                           \big(1 - e^{-\sigma_{n\chi}/\sigma_\text{sat}} \big) \,.
    \end{align}
    For the saturation cross-section we use the result of
    Ref.~\cite{Kouvaris:2007ay}
    \begin{align}
      \sigma_\text{sat} \simeq \frac{R^2}{0.45 N_b \xi} \,,
    \end{align}
    where $N_b$ is the number of baryons in the neutron star. 

  \item $B$ is a function accounting for the minimum energy loss necessary to
    capture a DM particle:
    \begin{align}
      B = \frac{6 v_\text{esc}^2}{\bar{v}^2} \frac{m_\chi m_n}{(m_\chi - m_n)^2}\,.
    \end{align}
    This function is smaller than one only when the DM mass is $\gtrsim \SI{E+6}{\GeV}$.
\end{itemize}
Using $C_\chi$ from \cref{eq:C-chi}, the number of DM particles captured will be
\begin{align}
  N_\chi = C_\chi t_\text{NS} \,.
\end{align}
Note that we here neglect the possibility of DM self-annihilation,
co-annihilation \cite{Bramante:2013nma} or semi-annihilation
\cite{DEramo:2010keq}, which would reduce $N_\chi$.

\bibliographystyle{JHEP}
\bibliography{refs}

\end{document}